\documentclass[10pt]{article}
\usepackage{graphicx}
\usepackage{capt-of}
\graphicspath{{c/}}
\DeclareGraphicsExtensions{.eps,.ps}
\begin{document}
\vspace*{2.5cm}
{\Large\bf The complete quantum collapse scenario of a 2 + 1 dust shell: 
Preliminary calculations}

\vspace*{0.8cm}
{\bf L. Ort{\'\i}z and M. P. Ryan, Jr.
\footnote{Instituto de Ciencias Nucleares, Universidad Nacional Aut\'onoma
de M\'exico, A. Postal 70-543, M\'exico 05249 D. F., Mexico}}

\vspace*{0.5cm}
\hspace*{1.5cm}\begin{minipage}[t]{9.5cm}\footnotesize{\it Abstract}

\hrulefill

If we consider the gravitational collapse of a material object to a black
hole, we would expect, for ranges of mass where a black hole would form, the 
following scenario.  A large enough object would collapse classically until
an event horizon forms, and to an external observer the object would be lost
from view.  However, once the horizon has formed, the black hole will begin to emit
Hawking radiation and the hole will lose mass and the horizon will shrink.
The final state of this process could be either a zero-mass ``black hole'' with consequent
information loss, or some sort of ``quantum remnant.''

     A complete investigation of this process would require: 1) A complete and
consistent theory of quantum gravity coupled to some kind of field that 
would provide the Hawking radiation \cite{hawk} (which could be the gravitational field itself ---
gravitons); 2) Some kind of definition of a ``horizon'' in this quantum gravity, and;
3) The calculational tools to achieve a description of the scenario. Lacking these, one may
resort to toy models to try to give some sort of a preliminary answer.  

In this paper we will consider the collapse of an infinitesimally thin
dust shell in 2 + 1 gravity, where an exact minisuperspace quantum solution 
exists, and try to make rough estimates of the collapse-Hawking radiation-remnant 
formation process.      

\hrulefill
\end{minipage}
\vspace{1cm}
\section{Introduction}

There has been some
interest over the years in the minisuperspace quantization of thin shells as
models of the full quantum collapse of more complicated objects.  One
important question that could be asked is whether enough information 
could be squeezed out of the grossly oversimplified model to mock up
a first look at the complete collapse scenario. In Ref. \cite{meleo} we used previous
work by Peleg and Steif \cite{stipel}, Israel \cite{Isra}, and Cris\' ostomo and Olea 
\cite{crisol} on the classical equations of motion of a dust shell to construct a 
quantum formulation of the collapse of such a shell.  

This quantum formulation was used to study the possibility of the formation
of a horizon, that is, whether some sort of ``quantum black hole,'' will form 
or whether a shell of non-interacting particles will simply collapse to a point
where the uncertainty principle will provide a ``repulsive force'' and the
shell will reexpand into the same universe.  
Simple considerations can answer this question up to a point.  It seems
obvious that for large enough masses one might expect a reasonably
peaked wave function centered on a radius near to but outside the
classical Schwarzschild radius of the shell would, as it moves toward
zero radius, maintain enough coherence to pass beyond the Schwarzschild
radius almost in its entirety and a ``black hole'' would form with high
probability.  For small masses one might expect that rapid spreading
would overwhelm the coherence of the wave function and the shell would
reexpand into our own universe with probability essentially (or exactly) one.

In full relativistic quantum gravity these simple ideas are fraught with
difficulties.  The most serious of these are:\\

1) The quantum shell problem becomes unphysical for masses much above the Planck
mass.  While the minisuperspace approach has frozen out all radiative modes
of the gravitational field, and one is insisting on a single-particle
interpretation of the shell problem, the wave function in the Schr\" odinger
picture can still become pathological at a point where one might expect
graviton production to begin.  H\'aj{\'\i}\v cek \cite{haj1}
has given a sufficient
but not necessary condition that shows that we might expect problems
above a couple of Planck masses.  In \cite{cruz}, a qualitative
argument was presented that used Compton wavelength considerations to give a
similar bound.

2) There are very serious technical problems in the formulation of the
problem.  Many authors have used the full ADM method to construct a 
Hamiltonian for the system,
where they have chosen an internal time for the system in order to have a
true Hamiltonian
and a Schr\" odinger equation that allows the study of the time evolution of
the quantum system.  The choice of an internal time leads to the problem
of quantum formulations that are not unitarily equivalent.  Another problem
with these Hamiltonian formulations is that they often require an {\it ad
hoc\/} choice of a Hamiltonian in terms of variables defined on the shell
itself.  Several of these Hamiltonians have been given by various authors
\cite{haj1}, \cite{hkk}.  A Hamiltonian due to
H\'aj{\'\i}\v cek and Kucha\v r \cite{hyk}
has the advantage of being defined by a coherent procedure with no {\it ad
hoc\/} choices, and is formulated in terms of foliations of spacetime
by timelike surfaces.  We will discuss this Hamiltonian in more detail
below.  All of these Hamiltonians have limits on the mass of the shell
of a few
Planck masses.  One formulation that does not seem to
have a mass limit is based
on the Wheeler-DeWitt equation corresponding to one of the Hamiltonians
in \cite{hkk}.

3) Many of the Hamiltonians are quite complicated and there is no real
chance of finding analytic solutions to the Schr\" odinger equations of
these models.  Numerical solutions have been presented by a group consisting
of A. Corichi, G. Cruz, A. Minzoni. M. Rosenbaum  and M. Ryan of the UNAM, N.
Smyth of the University of Edinburgh, and T. Vukasinac of the University
of Michoacan \cite{cruz}.\\

The idea of Ref. \cite{meleo} was to study the quantum collapse
problem in 2 + 1 gravity, where one can address some of the difficulties
mentioned above in a context where
some of the problems mentioned above do not exist. The quantum problem is
unambiguous for all masses, so there is no problem of wave function
pathologies.  As will be shown below, one possible Hamiltonian for this
problem has the form of that of a harmonic oscillator. This will allow us to find
simple analytic solutions that can be used to illustrate the development of the
quantum collapse of the shell, and allowed us to investigate the problem of
horizon formation.

The usual view of black hole evaporation and remnant formation has a vast literature.
A recent review article \cite{koch} cites a large number of authors.  This usual
view has three phases \cite{stromtom}\\

1) The {\it balding} phase, where the black hole radiates away all its multipole
moments, losing mass due to classical gravitational radiation.\\

2) The {\it evaporation} phase, where Hawking radiation of thermally distributed
quanta carries away mass until the Planck mass is reached.\\

3) The {\it Planck} phase, where quantum gravity is important and a remnant is
formed or all the mass is carried away.\\

We will not be interested in the balding phase, only in the circularly symmetric 
evaporation and Planck phases.  There are a number of arguments for remnant
formation (see \cite{koch} and references therein), one of which is similar
to what we will consider here, arguments based on the uncertainty principle.

Our idea is to investigate Hawking radiation in the quantum collapse of a
dust shell.  This is a minisuperspace model of quantum gravity associated
with collapse.  Our scenario is slight different from the usual view given
above, where the existence of a stable black hole created some time in
the past evaporates toward a stable remnant.  The back reaction of the
Hawking radiation causes the mass of the collapsing shell to change, 
affecting the shell evolution.  We will attempt to show that this effect
causes a shell that has collapsed below its (classical) horizon will
reappear with a reduced mass.

One possible way to calculate this scenario would be to define a 
``shell cloud'' from the probability density of shell radius, in the
same way that we define an ``electron cloud'' from the electron
probability density around a nucleus.  For the electron cloud we
can calculate its classical electromagnetic field, while for the
shell cloud we can calculate its classical metric.  The study of 
Hawking radiation in this background could be illuminating.

Since this idea is difficult to carry out, we will use a rough
approximation consisting of a classical shell that follows
the path of the expectation value of the quantum shell, and a model
Hawking radiation.  Since in Ref. \cite{meleo} we saw that a shell
wave packet collapses to a minimum radius and the reexpands due to
uncertainty principle effects, 
this scenario can have the
shell falling below its classical horizon, and the resulting black hole
begins to evaporate, and the shell, reappearing with mass loss as some sort
of ''dynamic remnant'' in the form of an expanding shell.    

The plan of the rest of the article is as follows.  Section 2 will be
a brief resum\'{e} of the literature on the general relativistic minisuperspace
problem.  Section 3 will consider the classical and quantum 2 + 1 problems and
discuss a model of horizon formation,
Section 4 will model the complete collapse-Hawking radiation-final state scenario, while
Section 5 will be conclusions and suggestions for future study.

\section{The collapse of thin shells in relativistic quantum gravity}

The study of the quantum collapse of dust shells is about a decade old \cite
{strom}.  The classical collapse problem is fairly straightforward, and
can in principle be solved exactly. One assumes a $\delta$-function massive
(or null) shell where Birkhoff's theorem tells us that outside the shell
the metric is Schwarzschild and the metric inside the shell is Minkowski.
The Israel junction conditions can be used to derive the equation for the
evolution of the shell in terms of intrinsic variables on the shell itself,
the proper time, $\tau$, of an observer riding on the shell and the curvature
radius, $R(\tau)$, of the shell that he would measure.  The equation for the
motion of the shell becomes \cite{Isra}
\begin{equation}
M = m\left \{ 1 + \left (\frac{dR}{d\tau}\right )^2 \right \}^{1/2} -
\frac{m^2}{2R}, \label{iseq}
\end{equation}
where $m$ is the rest mass of the shell (a constant of motion) and
$M$ is the Schwarzschild mass of the exterior metric.  It is straightforward
to define $x = R/m$, $\tau \rightarrow \tau/m$, $V \equiv dx/d\tau$ and $M/m$ as our
Hamiltonian.  Assuming $V = V(P)$ and solving
$\partial H/\partial P = V(P)$ for the ``momentum" $P$, we find that $V =
\sinh^{-1} (P)$ and our Hamiltonian becomes
\begin{equation}
H = \cosh P - \frac{m}{2x},  \label{hajham}
\end{equation}
a Hamiltonian
given by H\'aj{\'\i}\v cek \cite {haj1}. Unfortunately, this is not the only
Hamiltonian that gives the equation of motion (\ref{iseq}), and we are left
with the problem of defining an ``appropriate'' Hamiltonian for the
problem.  A number of Hamiltonians for different choices time, including
the time of an observer at the center of the shell where space is flat, and
a Wheeler-DeWitt equation identical to that for a relativistic charged
particle radially falling in a Coulomb potential are given
by H\'aj{\'\i}\v cek, Kay and Kucha\v r \cite{hkk}.

Kucha\v r and H\'aj{\'\i}\v cek \cite{hyk}, dissatisfied with such
{\it ad hoc\/} Hamiltonians,
have managed to construct a Hamiltonian for collapsing dust shells that
comes directly from an ADM reduction of the Hilbert-plus-matter action.
The problem with this approach is that, as spacetime quantities, the matter
variables are proportional to $\delta [R - R_0(\tau)]$, where $R_0 (\tau)$ is
the position of the shell as a function of shell proper time.  It is
virtually impossible to reduce the time derivatives of these delta functions
to reasonable variables in the matter Lagrangian that can give a shell
Hamiltonian that describes the motion purely in terms of canonical variables
on the shell.  Kucha\v r and H\'aj{\'\i}\v cek used an
ingenious method (following \cite{hajki}) based on the fact that
the ADM reduction by a $3 + 1$ foliation is not restricted to foliation
by spacelike surfaces, but works just as well for foliations by timelike
surfaces.  Unfortunately, this approach usually leads to an ill-posed problem, but
in the case of the shell it does not. Using this approach and a 
formulation of the dust fluid
velocity in terms of velocity potentials, they define a new Hamiltonian.
The cost of this consistent formulation is a very complicated Hamiltonian,
\begin{equation}
H= -\sqrt{2}R\left (1 - \frac{M}{R} - \sqrt{1 - \frac{2M}{R}}\cosh\frac{P}
{R}\right )^{1/2} \qquad R \geq 2M,            \label{karham1}
\end{equation}
\begin{equation}
H = -\sqrt{2}R\left (1 - \frac{M}{R} - \sqrt{\frac{2M}{R} - 1} \sinh
\frac{P}{R} \right )^{1/2} \qquad 0 \leq R \leq 2M.           \label{karham2}
\end{equation}

Almost all of the Hamiltonians that have been given
(except for the Wheeler-DeWitt equation of
 \cite{hkk}) seem to have mass limits beyond which the wave functions
become pathological.  These Hamiltonians are all so complicated that it
seems impossible to find analytic solutions to assist in the interpretation.  
In \cite{cruz}, Corichi et
al. have given a series of numerical solutions that give the evolution of
wave functions that are initially sharply peaked over a radius near the classical
horizon, $R = 2M$, for the $\cosh P$ Hamiltonian (\ref{hajham}) and an 
approximation to the
Hamiltonian given by Eqs. (\ref{karham1}-\ref{karham2}).  All of these solutions
show evolution of the peak toward $R = 0$ with a bounce caused by the
boundary conditions at $R = 0$, with the appearance of interference fringes as well as
a rapid spread of the wave packet.  The
Kucha\v r-H\'aj{\'\i}\v cek Hamiltonian has wave functions similar to those of
(\ref{hajham}), but with many rapid oscillations superimposed.

Since these Hamiltonians are self-adjoint, unitary evolution implies that
a peak formed from scattering states will always rebound to $R = \infty$.
One can ask whether this behavior means that all quantum collapse of this
sort implies a rebound into our own universe.  Since $\tau$ is proper time
on the shell and $R$ is also a shell variable, such questions can only
be answered by knowing the global quantum spacetime surrounding the shell.
In any case, the scenario of the shell observer is that he sees (begging
questions of quantum measurement and the reduction of
the wave packet) the shell
collapse to some point near $R = 0$, where uncertainty principle effects
change the classical equations of motion and the shell rebounds (actually,
a shell where the particles do not interact directly with one another
``passes through itself'' and reexpands, that is, each radially infalling
particle passes through $R = 0$ and the azimuthal angle $\theta$ jumps from
the initial $\theta_0$ to $\theta_0 + \pi$).  Even if the shell has
collapsed below its classical horizon, in finite proper time it will again
be above the horizon  and traveling toward $R = \infty$.  This
quasi-classical scenario is not surprising.  In proper time, a classical shell
that manages to avoid forming a curvature singularity at $R = 0$ would behave
in this way, but as $R$ becomes greater than $2M$ the shell would be in
a universe beyond our temporal infinity ($i_{+}$), or in ``another universe.''

This quasi-classical scenario is what one might expect to see for a large
mass where the quantum fluctuations would be small compared to the mass,
so the evolution of the wave packet would be coherent long enough for
the shell to collapse past its horizon and the shell would emerge from
the horizon into a new universe. However, if the wave packet spreads
sufficiently so that the width is greater than the classical horizon radius,
we can see that a horizon might never form, and the shell would reexpand
into our own universe.

The problem of horizon formation in the quantum system is very difficult.
Event horizons are global features and one has to try to define a global
feature in a fluctuating manifold.  Of course, this quantum manifold must
be constructed in terms of the full minisuperspace canonical
quantum gravity of the shell-metric system.
In the shell case we tend to use some kind of approximation to construct
the spacetime metric.  Kucha\v r \cite{ku2} argues that for the simple
shell minisuperspace we may just replace the the shell mass (Schwarzschild
mass) and the shell radius in the metric outside the shell by the
corresponding operators to make a ``metric operator.'' The problem with
this metric operator is that it is a function of shell proper time, and
studies of the metric close to the shell \cite{tatj} cannot tell us
whether a true event horizon (tied to observer time at infinity) forms.

Other approximations are under study \cite{us}.  The simplest calculation
is to calculate $<\hat R(\tau)>$ and the uncertainty $\Delta R = \sqrt{<(\hat R - <\hat R>)^2>}$
and check whether $\Delta R$ becomes very large as $<\hat R>$ becomes small so that
$<\hat R>$ does not fall below the classical horizon and $\Delta R$ is larger than
 the classical horizon, which
can be taken as an indication of the non-formation of a horizon.  In 
\cite{us}, numerical evaluations of these two quantities will be presented
for the Hamiltonians (\ref{hajham}) and (\ref{karham1}-\ref{karham2}), and
they
suggest that no horizon forms for small masses.  Another possibility (to
be considered in \cite{us}) would be to take M$|\psi (R, \tau)|^2$ to be a
classical density $\rho (R, \tau)$ and calculate the classical metric due
to a classical fluid with this mass distribution and see whether a horizon
forms. Note that $M$ should be either the rest mass or the Schwarzschild
mass.  It is not yet clear which.  There are
technical problems with this calculation.  We have to
calculate the metric from a density that is given in terms of a solution
of the Schr\" odinger equation for our Hamiltonian, and there is no
guarantee that this density can be made to obey the equation
$T^{\mu \nu}_{\, \, \, \, \, ;\nu} = 0$ for our fluid. Another possibility
considered in
\cite{us} is that of the ``metric operator'' mentioned above, which was
extended to the whole
manifold outside the shell and used to define a ``quantum'' stress-energy
tensor.

The rotationless 2 + 1 problem has some advantages over the 3 + 1 problem.
The Hamiltonian can be constructed fairly easily, and, as will be shown,
has the form of a harmonic oscillator. The Schr\" odinger
equation for this Hamiltonian has well-known analytic solutions.  The
expectation value of $R$ and its uncertainty can, in principle, be
calculated analytically.  The analogue of the other calculation using $\rho
(R, \tau)$ is much simpler than in the 3 + 1 case. The classical horizon is
easily
found.  In the next section these ideas will be considered.
 
\section{The classical and quantum collapse of shells in 2 + 1 gravity}

The first element we need for this problem is an equation for the radius of
the shell.  This problem has been studied in detail by Peleg and Steif 
\cite{stipel}, using the 2 + 1 version of the original  formulation of 
Israel \cite{Isra}, and Cris\' ostomo and Olea \cite{crisol}, using canonical methods. 

Israel
studied the collapse of a shell in 3 + 1 gravity, represented by a 
delta-function sphere of dust of
radius $R(\tau)$.  In 2 + 1 gravity the shell is a
circle, i.e.
a ring of matter, also of radius $R(\tau)$. The metric of spacetime will be 
written in circular coordinates, where flat space is represented by the metric
\begin{equation}
ds^2 = -dt^2 + dr^2 + r^2 d\theta^2.
\end{equation}

The equation for a circle $R(\tau)$ is
\begin{equation}
^{(3)}r = R(\tau), \qquad ^{(3)}\theta = \theta, \qquad ^{(3)}t = t(\tau).
\end{equation}
We will use the notation $i, j = 1, 2, 3$ and $A, B = 1, 2$.  We will now need 
a set of coordinates $\xi_A$ on the circle, which we will take to be $\xi_A =
(\tau, \theta)$.

Off the ring of matter, the 2 + 1 version of Birkhoff's theorem says
that the three-dimensional metric is a static or stationary solution to 
\begin{equation}
R_{ij} - \frac{1}{2}g_{ij} R = -\Lambda g_{ij},
\end{equation} 
(where we have to have a cosmological constant $\Lambda$ to avoid locally completely 
flat solutions) both
inside and outside the circle. Here, as in Ref. \cite{meleo}, we will study
the static case,
where the matter has no angular momentum.  In this case the metric
has the form
\begin{equation}
ds^2 = -f(r)dt^2 + \frac{1}{f(r)}dr^2 + r^2 d\theta^2,
\end{equation}
and the well-known solution is
\begin{equation}
f = B - \Lambda r^2,
\end{equation}
$B$ a constant.  Since $\Lambda$ has units of inverse length, we will
write, as is common, $\Lambda = \pm 1/\ell^2$.
The final form of the static, circularly-symmetric metric is
\begin{equation}
ds^2 = - \left( -M \mp \frac{r^2}{\ell^2}\right ) dt^2 + \frac{1}{\left ( -M \mp \frac{r^2}{\ell^2}\right )}
dr^2 + r^2 d\theta^2, \label{metrc}
\end{equation}
where we have taken $B$, following Ba\~nados, Teitelboim and  Zanelli (BTZ) \cite{BTZ}, to be $-M$,
M a Schwarzschild mass.  We will take 
the metric inside the ring to be the 2 + 1 AdS metric where $M = -1$ 
\cite{BTZ2}, \cite{carl}.   As in the 3 + 1 case, we
expect that outside the ring we will have a black hole metric with some
``Schwarzschild mass" $M$, that is,
\begin{equation}
ds^2 = - \left(-M + \frac{r^2}{\ell^2}\right ) dt^2 + \frac{1}{\left (-M + \frac{r^2}{\ell^2}\right )}
dr^2 + r^2 d\theta^2, \label{metfin}
\end{equation}
and inside the ring, 
\begin{equation}
ds^2 = - \left( 1 + \frac{r^2}{\ell^2}\right ) dt^2 + \frac{1}{\left ( 1 + \frac{r^2}{\ell^2}\right )}
dr^2 + r^2 d\theta^2.
\end{equation}

With these preliminaries, in Ref. \cite{meleo}
the equation of motion of the shell was calculated using the
2 +  1 analogue of the Israel formulation of the equation of
motion of a shell in 3 + 1 gravity, which was similar to that
used by Peleg and Steif \cite{stipel}.  

In \cite{meleo} we used the jump in the Einstein equations between the interior and
exterior of the shell, and
the Lanczos relation,
\begin{equation}
\gamma_{AB} - g_{AB}\gamma = 8\pi S_{AB}, \label{lanc}
\end{equation}
where $\gamma_{AB}$  is  the jump of $K_{AB}$, the extrinsic curvature of the shell as
seen from  the inside and the outside of the shell, the $g_{AB}$ are the components of 
the induced metric on the surface in
terms of $\tau$ and $\theta$,
$\gamma = g^{AB}\gamma_{AB}$, and $S_{AB}$ is the surface stress-energy
tensor, (In our case, we will
be interested in a dust shell, so we will take $S_{AB} = \sigma u_A u_B$,
$\sigma$ the rest mass density
of the ring, where $\sigma = m/2\pi R$, $m$ the total rest mass of the shell.)
to find the classical equation of motion for the shell,
\begin{equation}
\ddot R = -R/\ell^2.\label {eq2}
\end{equation}

Equation (\ref{eq2}) has a first
integral,
\begin{equation}
E = \frac{1}{2} \dot R^2 + \frac{R^2}{2\ell^2},\label{eeq}
\end{equation}
and, using the Lanczos equation directly,
we find
\begin{equation}
\sqrt{1 + 2E} - \sqrt{-M + 2E} = 4m,
\end{equation}
which can be solved for E as
\begin{equation}
E = \left (\frac{M}{32m} + \frac{1}{32m} + \frac{m}{2}\right )^2 - \frac{1}{2}.
\label{eeq2}
\end{equation}

Equation (\ref{eeq}) was used to construct a Hamiltonian
formulation of the problem.
This equation is simply the energy equation for a harmonic oscillator,
so it is obvious that
if we take $E$ as our Hamiltonian we can define a
momentum $P_R$ as $\dot R$, and we have
\begin{equation}
H = \frac{1}{2}P_R^2 + \frac{R^2}{2\ell^2}.
\end{equation}
Hamilton's equations for this Hamiltonian are equivalent to the equation
of motion $\ddot R  = -R/\ell^2$.

Once we have a Hamiltonian in terms of the ring variables $R$ and $\tau$,
it is possible to construct a Schr\" odinger equation for the problem.
If we write the classical relation
\begin{equation}
\left (\frac{M}{8m} + \frac{1}{8m} + 2m\right )^2 - 1 = \dot R^2 + \frac{R^2}{\ell^2},
\label{hameq1}
\end{equation}
on the right-hand-side $M$ and $m$ are dimensionless, so $M \equiv M/M_{\rm Pl}$ and $m 
\equiv m/M_{\rm Pl}$, where $M_{\rm Pl}$ is the Planck mass, $M_{\rm Pl} = c^2/G^{(2)}$
(we will later need the Planck length, $L_{\rm Pl} = \hbar G^{(2)}/c^3$), where $G^{(2)}$
is the two-dimensional Newton's constant with units $[L^2]/[M][T^2]$.

In conventional units, our equation for $R$ becomes
\begin{equation}
\left (\frac{M}{8m} + \frac{M_{\rm Pl}}{8m} + \frac{2m}{M_{\rm Pl}}\right )^2 - 1 = \frac{1}{c^2}
\left (\frac{dR}{d\tau} \right )^2 +
\frac{R^2}{\ell^2}.
\end{equation}
Our ``Hamiltonian'' should have the units of energy for a conventional
quantum Hamiltonian, so
we should multiply Eq. (\ref{hameq1}) above by $M_{\rm Pl} c^2$, our
$R$-momentum $P_R$ becomes
$M_{\rm Pl} \dot R$ , and with
\begin{equation}
H = \frac{M_{\rm Pl}c^2}{2} \left (\frac{M}{8m} ^2 + \frac{M_{\rm Pl}}{8m} + 
\frac{2m}{M_{\rm Pl}}\right)^2 - \frac{M_{\rm Pl}c^2}{2}, \label{hamval}
\end{equation}
and
\begin{equation}
H = \frac{P^2_R}{2M_{\rm Pl}} + \frac{M_{\rm Pl}}{2}\omega_0^2 R^2,
\end{equation}
where $\omega_0 = c/\ell$.

In \cite{meleo} we quantized this system with $H$ replaced by an operator $\hat H$.
If we look at (\ref{hamval}), we see that the right-hand-side must become an
operator, so the left-hand-side must also be an operator. We also took 
$M$ to be a $q$-number, and $m$ and $M_{\rm Pl}$ $c$-numbers.
Of course, there is no special reason to make this choice, but we do so to
make contact with previous work (see, for example, \cite{ku2}).  If we
consider Eq. (\ref{hajham}), this choice is motivated by the fact that
the Schr\"odinger equation for (\ref{hajham}) is similar to the hydrogen atom
Schr\"odinger equation, with $m$ playing the role of $e^2$, and it is usual in the
hydrogen atom to take $e^2$ as a $c$-number rather than a $q$-number.

We can now take the wave function of the system to be $\tilde \psi =
\psi_M \psi (R, \tau)$, with $\psi_M (M)$ an approximate eigenstate of
$\hat M$ with eigenvalue $M_0$.  An exact eigenstate of $\hat M$ of this
type would be $\delta (M - M_0)$, but to avoid problems with the integral of
the square of a delta function we will assume that $\psi_M$ is an
extremely sharply peaked wave function centered on $M = M_0$.  In this case,
our Schr\" odinger equation becomes ($\hat M^2 \psi_M \approx M_0^2 \psi_M$,
and realizing $\hat P_R$ as $-i\hbar \partial/\partial R$)
\begin{eqnarray}
\left [  \frac{M_{\rm Pl}c^2}{2} \left (\frac{M_0}{8m}+
\frac{M_{\rm Pl}}{8m} + 
\frac{2m}{M_{\rm Pl}}\right)^2 -
\frac{M_{\rm Pl}c^2}{2}\right ] = \nonumber \\
= i\hbar \frac{\partial \psi(R, \tau)}{\partial \tau} = -\frac{\hbar^2}
{2M_{\rm Pl}} \frac{\partial^2\psi (R, \tau)}{\partial R^2} +
\frac{M_{\rm Pl}}{2}\omega_0^2 R^2 \psi (R, \tau).
\label{schreq}
\end{eqnarray}
One curious fact about this equation is that, if we define $\psi =
\exp(-iE\tau/\hbar)\psi(R)$, then the energy eigenvalues are
\begin{equation}
E_n = (n + \frac{1}{2})\hbar \omega_0 = (n + \frac{1}{2})
\frac{\hbar c}{\ell},
\end{equation}
and since $E$ is given by the first line of (\ref{schreq}), there is a
discrete relation between the Schwarzschild mass, $M_0$, and the rest mass,
$m$,
\begin{equation}
\left (\frac{M_0}{8m} + 
\frac{M_{\rm Pl}}{8m} + 
\frac{2m}{M_{\rm Pl}}\right)^2  - 1 =
(2n + 1)\frac{\hbar G}{c^3 \ell} = (2n + 1)\frac{L_{\rm Pl}}{\ell}.
\end{equation}
We will now return to units where $G = c = \hbar = 1$, $M_0$ now meaning
$M_0/M_{\rm Pl}$, $m$ now meaning $m/M_{\rm Pl}$, and $\ell$ meaning
$\ell/L_{\rm Pl}$. If we solve for $M_0$ in terms of $m$ and $\ell$, we find
\begin{equation}
M_0 = 16m^2\left [\sqrt{\frac{1}{4m^2} + \frac{1}{2m^2}\left[
\frac{1}{\ell}\left (n + \frac{1}{2}\right )\right ]} - 1\right ] - 1.\label{meq}
\end{equation}

For a moment we will return to conventional units, and define
the following dimensionless variables.  We define $y = R/\sqrt{\ell
L_{\rm Pl}}$ and
$T = c\tau/\ell$. The Schr\" odinger equation now becomes
\begin{equation}
i\frac{\partial \psi (y, T)}{\partial T} = -\frac{\partial^2 \psi (y, T)}
{2\partial y^2} + \frac{y^2}{2} \psi(y, T)
\end{equation}
which has eigensolutions
\begin{equation}
\psi = e^{-i(n + \frac{1}{2})T}\frac{1}{\sqrt{2^n n!}(\pi)^{1/4}}
e^{-y^2/2}H_n (y),
\end{equation}
$H_n$ Hermite polynomials.

In Ref. \cite{meleo} we studied
the evolution of a wave packet sharply peaked around a value of $y = y_0$,
a point some distance outside the point where a classical horizon would form if the
radius of the shell were to fall below $R_H = \ell \sqrt{M_0}$ and
followed its movement as the packet fell toward $R = 0$.
In the 2 + 1 case it is possible to give an exact analytic
expression for the wave packet as a coherent harmonic-oscillator state.
However, even though the eigensolutions are nothing but harmonic oscillator wave
functions, the radial variable $R$ cannot be negative.  In order to keep
this from happening we will took the potential to be that of a half
oscillator with an infinitely hard wall at $R = 0$ This potential is
shown in Figure 1

\vspace{1cm}
\centerline{\scalebox{.4}{\includegraphics{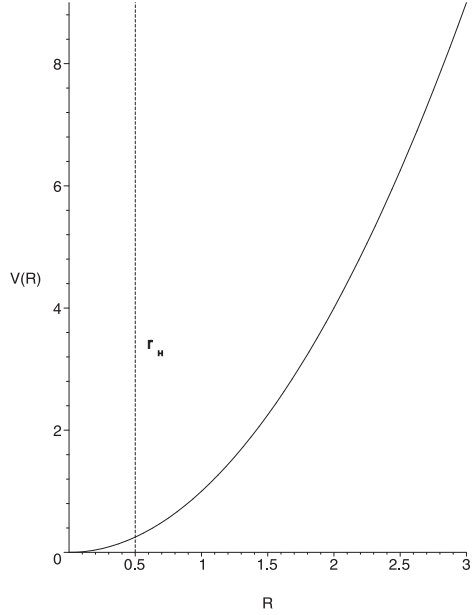}}}
\captionof{figure}{\small The harmonic oscillator potential for our problem with $M_{\rm Pl} \omega_0^2/2 = 1$.  The dashed line shows a typical position of the classical horizon radius, $r_H$.}
\vspace{0.5cm}

\noindent This meant that $\psi (0, \tau)$ was always
zero.  This boundary condition can be enforced by only expanding in odd-$n$
harmonic oscillator eigenfunctions. 

In \cite{meleo} we began with a difference
of two Gaussian states, one peaked around $y = +y_0$ and the other around
$y = -y_0$, so that their sum at $y = 0$ was zero.
This state is
\begin{equation}
\psi (y, 0) = \alpha [e^{-\frac{1}{2}(y - y_0)^2} - e^{-\frac{1}{2}(y +
y_0)^2}],\label{init}
\end{equation}
only valid for $y > 0$, with $\alpha$ a normalization constant.  Since this
is a sum of two Gaussian states, we can use standard techniques to construct
the difference between two coherent Gaussian states with (\ref{init}) as an
initial condition.  The result is
\begin{eqnarray}
\psi(y,\tau) =  \alpha e^{-i\omega_0 \tau/2} e^{i y_0^2 \sin 2\omega_0
\tau/4}\times \\ \nonumber \label{cohstat}
 [\exp(-\frac{1}{2}\{ (y - y_0\cos \omega_0 \tau )^2\})
 \exp (-iy y_0 \sin \omega_0 \tau )\\ \nonumber 
 - \exp(-\frac{1}{2}\{ (y + y_0\cos \omega_0 \tau)^2 \} ) \exp (iy y_0
 \sin \omega_0 \tau) ], 
\end{eqnarray}
which is zero at $y = 0$ for all $\tau$.  The normalization $\alpha$ is
easily found to be
$\alpha^2 = [\sqrt{\ell}\sqrt{\pi}(1 - e^{-y_0^2})]^{-1}$.

We connect the variables that describe the wave function with
the radius of the
classical horizon, $r_H$, shown in Fig. 1.
The
unnormalized probability density, $\rho \equiv \psi^{*} \psi/\alpha^2$ is
\begin{equation}
\rho = e^{-(w - \lambda \cos T)^2} + e^{-(w + \lambda \cos T)^2}
-2e^{-w^2} e^{-\lambda^2 \cos^2 T} \cos(2\lambda w \sin T), \label{wpsi}
\end{equation}
(where $w = y/\sqrt{\ell}\sqrt{M}$). 
Figures showing the evolution of this probability density are given in \cite{meleo}.
This evolution begins with a Gaussian packet at $T = 0$ that collapses toward $y = 0$,
developing interference fringes as it reaches a minimum for some
$y < y_0$, then rebounds toward $y = y_0$ again.
This pattern then repeats forever.  We will only be interested in one cycle of this pattern.
      
\subsection{The formation of a horizon}

In previous articles the quantization of the shell collapse was used to
study the possibility of the formation of
a horizon in the quantum collapse.  As mentioned in Sec. 2, this concept has
many difficulties.  An event horizon is a global construct and it has no
local definition.  This means that in quantum gravity one would have to
return to the starting point and try to define what a ``quantum horizon''
might be.  Once this definition has been decided upon, one must try to find
out if some collapse process will result in the formation of such a horizon,
with the result being a probability of
horizon formation.  Theories of quantum gravity in their present state are
far from being able to give us this result,
so, in shell collapse, some articles \cite{cruz}, \cite{us}, \cite{vinart}
have tried to give an estimate of horizon formation by finding out if a
sharply peaked wave packet, during its collapse toward $R = 0$, will fall,
in some sense, below the classical horizon radius, $r_H$.  In some sense,
because the packet will usually spread and basically will never lie entirely
below $R = r_H$. One could use the integral of $\psi^{*} \psi$ from $R = 0$
to $R = r_H$, which is a number less than one, as the probability of horizon
formation.  Another possibility would be to use the operator $\hat M$ in the
expression $\hat R_H = \ell\sqrt{2\hat M - 1}$ to define a
``horizon operator'' that could be used to define the probability of horizon
formation.  In previous work it was decided to use $<\hat R>$ as a function
of $\tau$, a quantity that falls from $R_0$ (the $R$ associated with $y_0$)
to a minimum and then begin to
increase again. If this minimum is below the classical horizon, one
can say that a horizon forms and if not, not.  This is a yes-no answer
instead of a probability, but it is
a quick estimate.  We will use this concept below.

It is not difficult to calculate $<\hat R>(\tau)$ from the wave packet given
in (\ref{cohstat}), but the result is a complicated function that contains
the error function and Dawson's function, so trying to find the minimum of
$<\hat R>(\tau)$ by finding the point where $d<\hat R>/d\tau = 0$ requires
the solution of a transcendental algebraic equation, making it difficult to
give an analytic expression for the point where a horizon would form.
To avoid this problem, in Ref. \cite{meleo} we used 
the fact that at $T = \pi/2$, the point
where the packet begins to rebound, the peak nearest $R = 0$ is high and
narrow.  We used the position of this peak as a parameter to tell us
whether a horizon would form or not.  If the position of the peak is below $r_H$,
a horizon forms, and if not, not. In previous work it was found that for
large masses the shell collapse was so rapid that the wave packet fell below
$r_H$ so quickly that quantum mechanics did not allow it to rebound before
that point, while for small masses the rebound occurred for $R > r_H$.  We
were able to give a range of masses where no horizon would form.

In the present article, we will study the behavior of $<\hat R>(\tau)$, but 
we will use the threshold mass where a horizon will form from \cite{meleo} 
in order to estimate the point where a horizon forms.  We will not give any details
of the calculation in \cite{meleo}, only the results.

We can define several purely numerical quantities that will be used to
explain the mass limits.  We will key off $r_H = \ell \sqrt{M}$, the value of $R$ of the
shell as it passes its classical horizon.  We will also make use of the
variable $y$.  The peak of the initial wave function is at $y_0$, and
using the fact that the value of $y$ corresponding to the classical horizon
is $y_H = \sqrt{\ell} \sqrt{M}$, we can define $y_0 = \lambda y_H$, $\lambda >
1$.  The point where the largest peak reaches its minimum at $y_c = \gamma y_H$.

Studying the motion of the wave packet, it was possible to find an analytic
relation between $\lambda$ and $\gamma$ in terms of $M_0$, where, for 
a given $\lambda$, $M_0$ is a monotonically decreasing function of $\gamma$ (and $\ell$).
Since $\gamma$ must be less than $\lambda$ for collapse to make sense, we have
a minimum of $M_0$ as a function of $\ell$.  Another minimum can be found by
studying $H$ as a function of $M_0$, $m$ and $\ell$.  If we take $<2\hat H>$
equal to $\left (\frac{M_0}{8m} + 
\frac{M_{\rm Pl}}{8m} +  \frac{2m}{M_{\rm Pl}}\right)^2  - 1$, and calculate
$<2\hat H>$ as a function of $M_0$, $m$ and $\ell$, we find a real 
solution for $M_0$ only for $m$ greater than a minimum value.  Taking $m$
to be this minimum, we find another minimum for $M_0$.  Equating these two
minima, we find a numerical value for $\ell$, $\ell \approx 0.18$.  For this
value of $\ell$, we find that for $M_0 < 1.48$ there is no real solution for
$M_0$, and for 
\begin{equation}
1.48 < M_0 < 3.32,
\end{equation}
no horizon forms, while for $M_0 > 3.32$ one does.
                                             
Of course, as has been mentioned above, the rebound of the wave packet,
once it has passed the horizon does not mean that the shell returns through
the horizon into the same spacetime where it began. In spite of the fact
that in terms of proper time the shell exits the horizon in a finite time,
in terms of the time, $t$, of an observer at infinity this exit occurs at
time {\it after} $t = \infty$, or into ``another universe.''

\section{The complete collapse--Hawking radiation--final state scenario}

As we mentioned in Sec. 1, we want to consider the complete evolution of a
collapsing quantum shell.  Without a complete, consistent quantum theory of 
gravity, we have to appeal to some sort of approximation.  One possible 
approximation that might be the closest we can come to a real theory would 
be to define a ``shell cloud'' in the same way we define an ``electron cloud''
when we consider such concepts as electron shielding of the charge of a
nucleus.  If we consider a hydrogen atom with the electron in some quantum
state, $\psi({\bf x}, t)$, we can define an electric charge density by
$\rho = e\psi^{*} \psi$, and then calculate the electromagnetic field
due to this ``electron cloud.''  We can do the same with the shell, that is,
define a classical mass density as something like $\rho = M_0 |\psi (R, \tau)|^2_{R 
\rightarrow r}$, and try to calculate a classical metric $g_{ij} (r, \tau)$ from
this mass density.     

Of course, $\rho(r,\tau)$ is not enough to solve the 2 + 1 Einstein equations,
we need a real stress-energy tensor, $T^{ij}$.  If we can define such a $T^{ij}$, we
can write the spherically symmetric metric as
\begin{equation}
ds^2 = -e^{\nu(r, t)}dt^2 + e^{\lambda(r, t)}dr^2 + r^2 d\theta^2,
\end{equation}
which, if $T^{ij} = T^{ij}(r,t)$, we have(see Synge \cite{sing})
\begin{equation}
e^{-\lambda} = -\Lambda r^2 + 16\pi \int^r_0 r T^0_0 dr,
\end{equation}
\begin{equation}
\nu = -\lambda - 16\pi \int^r_0 r e^{\lambda} (T^0_0 - T^r_r)dr.
\end{equation}
It is easy to relate $T^0_0$ to the mass density $M\sigma \psi^* \psi$,
where $\sigma$ is a numerical factor to take into account the fact that 
we want $e^{-\lambda}$ to be $-\Lambda r^2 - M$ for large $r$, but 
$T^r_r$ is undefined. We also have to study the rest of the components
of $T^i_j$ for possible singular behavior.  We can show, as in the 3 + 1 case,
that $T^{ij}$ should diagonal except for $T^{0r}$.  This means that we need
only find $T^{rr}$ and $T^{\theta \theta}$.  The Bianchi identities give
\begin{equation}
T^r_0 = -\frac{1}{r} \int^r_0 r\frac{\partial (T^0_0)}{\partial t} dr,
\end{equation}
\begin{equation}
T^{\theta}_{\theta} = rT^0_0 + r\frac{\partial (T^0_r)}{\partial t},
\end{equation}
so, in reality, we only need to define $T^r_r$ as long as $T^0_0$ can be calculated from
from $M\sigma \psi^* \psi$.  We can consider a fluid stress energy of the type
\begin{equation}
T^i_j = (\rho + p)u^i u_j + p\delta^i_j, \qquad u^{\theta} = 0.
\end{equation}
From $u^i u_i = -1 = -e^{\nu}(u^0)^2 + e^{\lambda} (u^r)^2$, and 
defining $e^{\lambda} (u^r)^2 \equiv U^2$, we have
\begin{equation}
T^0_0 = -(1 + U^2)\rho - U^2 p,
\end{equation}
\begin{equation}
T^r_r = U^2 \rho + (1 + U^2) p,
\end{equation}
and
\begin{equation}
T^0_0 - T^r_r = -(1 + 2U^2)(\rho + p),
\end{equation}
and
\begin{equation}
T^{\theta}_{\theta} = p.
\end{equation}
We can have $T^0_0 - T^r_r = 0$ if $p = -\rho$ and
$T^0_0 = -\rho$.  If we have $\rho = M\sigma \psi^* \psi$, and since we have taken 
the Schr\" odinger equation to have the form of Eq. (\ref{schreq}), what we call $\psi$ is actually
$\sqrt{r} \psi$, we will have (taking $\sigma = 1/16\pi$ in order to make $e^{-\lambda}$ 
 go to $-\Lambda r^2 - M $as $r \rightarrow \infty$)
\begin{equation}
e^{-\lambda} = -\Lambda r^2 - M\int^r_0 \psi^* \psi (r, \tau)dr,
\end{equation}
and $\nu = -\lambda$.

We can calculate $T^r_0$ and $T^{\theta}_{\theta}$, which in this particular case 
($T^0_0 - T^r_r = 0$), have singular points.  A better choice of $U$ would make $T^i_j$
well behaved, but we have an almost infinite choice of $U$, although tying $U$ to
the quantum mechanical current associated with our Schr\" odinger equation (\ref{schreq}) would
be best.

Once we have a reasonable metric (taking into account the caveats we have mentioned)
we could, in principle calculate the Hawking radiation in this background, as well as the
effect of its back reaction on the quantum evolution of the shell.  This is still a massive
undertaking, involving difficult numerical analysis.

Our plan in this article is to attempt a simplified analysis, where we can use purely 
analytic constructs (with the exception of a numerical solution to a simple first-order 
ODE) to model the complete collapse scenario.

The first element we need is some quantity to represent the shell evolution.  We will
consider the expectation value of $\hat R$, $<\hat R> (\tau)$, to represent a classical 
shell evolving with this $R(\tau)$.  We have
\begin{eqnarray}
<\hat R>(\tau) = \frac{\lambda \ell \sqrt{M}}{1 - e^{-\lambda^2 \ell M}} 
[\cos T {\rm erf}\, (\lambda \sqrt{\ell} \sqrt{M} \cos T) \nonumber \\
 - \sin T e^{-\lambda^2 
\ell M} i{\rm erf}\, (i\lambda \sqrt{\ell} \sqrt{M} \sin T)].
\end{eqnarray}
Figure 2 shows $<\hat R>(\tau)$ for $\lambda = 3$, $\ell = 0.18$, and $M = 5$.

This expectation value begins at $<\hat R> \approx 1.20$, falls to a minimum and rises again
to the original value.  In Fig. 2, the value of the classical horizon $r_H \approx 0.4$,  is shown as a
dashed line, and $<\hat R>(\tau)$ dips below this value, and later (after a relatively
short proper time interval) rises above $r_H$ again.  Of course, the amount of time 
$\Delta t$ measured by an observer far from the black hole between the moment when 
$<\hat R>(\tau)$ dips below the classical horizon and the moment when it reappears
is greater than infinity.\\

The next element needed for the complete scenario is Hawking radiation.
While Hawking radiation has been calculated for 2 + 1 gravity \cite{lifort}
\cite{Hlee}, in the spirit of 
our simplified calculations, we will use a quick estimate of the average 
energy of particles emitted in Hawking radiation and of $T_H$, the Hawking
temperature.  In the Appendix we calculate this energy in 3 + 1 gravity by
assuming that pair production occurs in regions of the size of the Compton
wavelength of the particle produced, $\lambda_c$, and that if a pair is
produced in a small region between $r_H$ and $\lambda_c$, one of the pair
may fall into the black hole, and the other appears as a real particle 
with velocity $c$.  

\vspace{1cm}
\centerline{\scalebox{.4}{\includegraphics{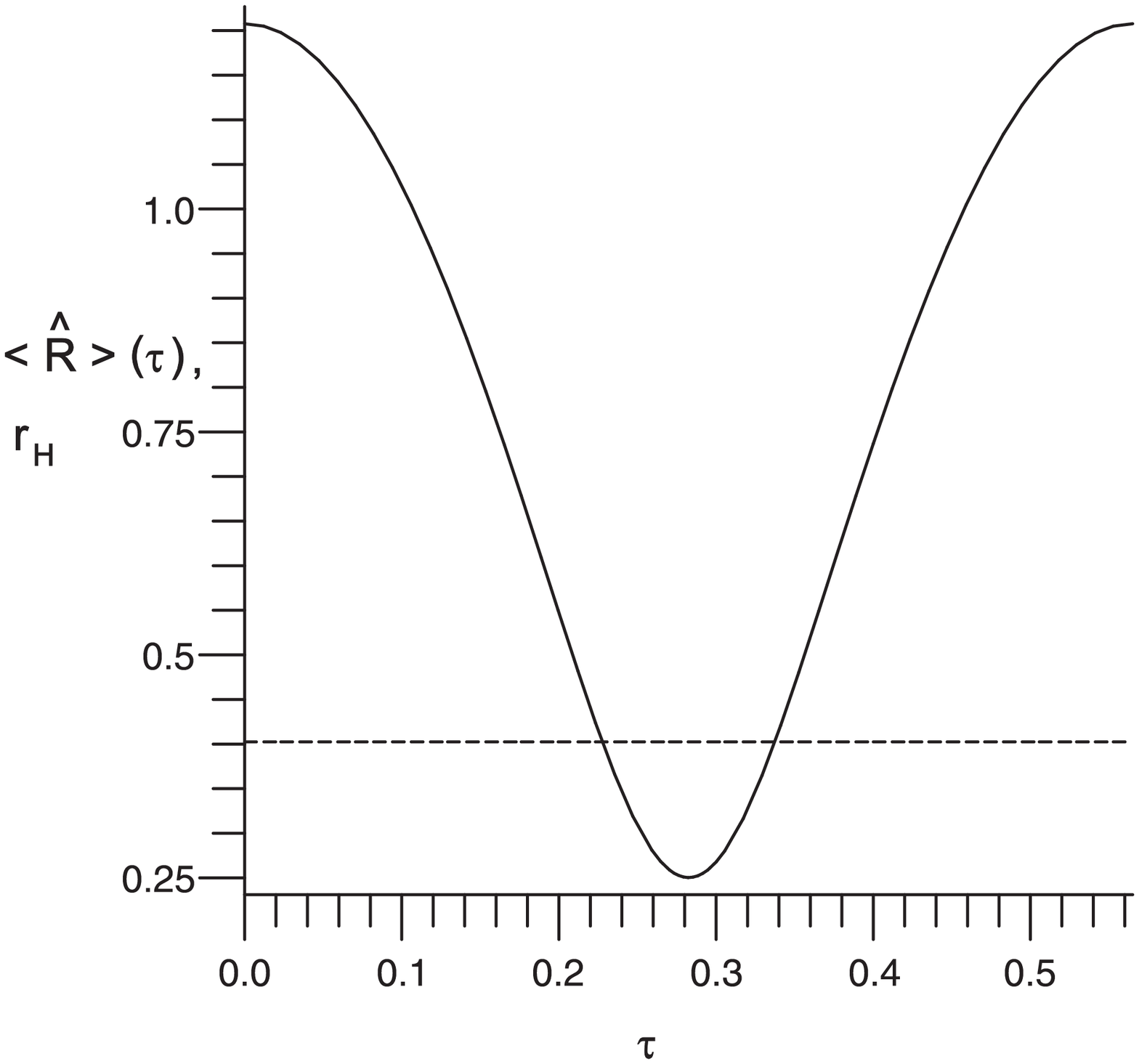}}}
\captionof{figure}{\small The quantity $<\hat R>(\tau)$ as a function of $\tau$ from $3r_H$ until it 
returns to $3r_H$ again for $M = 5$.  The dashed line shows the position of the 
horizon $r_H = \ell \sqrt{M}$.}
\vspace{0.5cm}

\noindent A simple Newtonian calculation of the energy at infinity
of this particle, assuming that it is created at $r_H + \lambda_c/2$ and 
moves radially toward $r = \infty$, gives an energy that is almost equal
to the average particle energy in Hawking radiation.

In 2 + 1 gravity there are several difficulties.  One is that there is
no Newtonian limit to 2 + 1 gravity, and another is that there is a
cosmological constant.  Since, in the Appendix, we use Newtonian
calculations, we can do the same here (using two-dimensional Newtonian
gravity).  For the cosmological constant we can use the Newtonian 
cosmological constant introduced by Seeliger and Neumann \cite{seenue} at the
end of the nineteenth century.

The two-dimensional Newtonian equation of motion for the radial
motion of a particle of mass $m$ with a cosmological constant in
the spherical field of a point mass $M$ is
\begin{equation}
m\frac{d^2 r}{dt^2} = - \frac{G^{(2)} Mm}{r} - \frac{mc^2}{\ell^2}r,
\end{equation}
where $c^2/\ell^2$ is the Newtonian cosmological constant, which has
a first integral
\begin{equation}
E = \frac{m}{2} \left (\frac{dr}{dt}\right )^2 + G^{(2)} Mm \ln r + 
\frac{mc^2}{2\ell^2}r^2.
\end{equation}
From Figure 3 we have $r_H = \ell  \sqrt{G^{(2)} M/c^2}$, and $\lambda_c =
\hbar/mc$, 

\vspace{1cm}
\centerline{\scalebox{.3}{\includegraphics{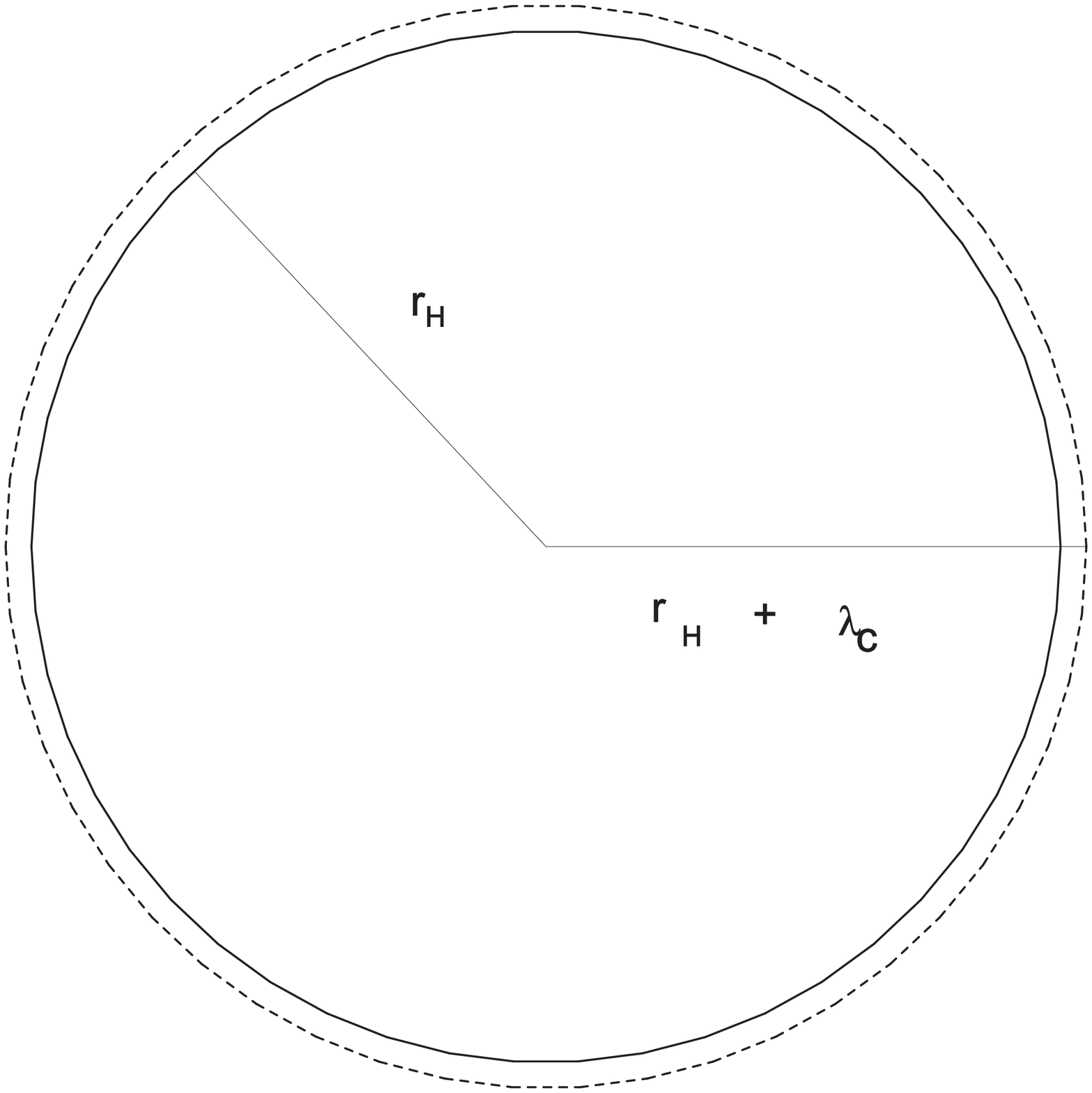}}}
\captionof{figure}{\small The circles showing the horizon $r_H$ and the thin region between $r_H$ and $r_H + \lambda_c$,
$\lambda_c$ the Compton wavelength of a particle of mass $m$.}
\vspace{0.5cm}

\noindent and the energy of a particle moving outward at velocity
$c$ at a position $r_H + \lambda_c/2$ is
\begin{equation}
E = \frac{mc^2}{2} + G^{(2)}Mm \ln \left (\ell \sqrt{\frac {G^{(2)}M}{c^2}} 
+ \frac {\lambda_c}{2}\right ) + \frac{1}{\ell^2} \left (\frac{\ell \sqrt{G^{(2)} 
M}}{c^2} + \frac{\lambda_c}{2}\right )^2 mc^2.
\end{equation}
For $\lambda_c << r_H$,
\begin{equation}
E \approx \frac{mc^2}{2} + G^{(2)} Mm \left [1 + \ln \left 
(\ell \sqrt{\frac{G^{(2}) M}{c^2}} \right ) \right ]  + \frac{3}{4} 
\hbar \frac{\sqrt{G^{(2)} M}}{\ell}.
\end{equation}

The first term is not zero as in the 3 + 1 problem, but the last term
can be taken to be the average energy of Hawking radiation particles.
This quantity is close to the average energy calculated in \cite{lifort} 
\cite{Hlee}, $E =
\hbar \sqrt{G^{(2)} M}/2\pi \ell$, and for our approximation we will
take it as the average energy.  We can now define the Hawking temperature
($k_B$ the Boltzmann constant),
\begin{equation}
T_H = \frac{3\hbar}{4k_B} \frac{\sqrt{G^{(2)} M}}{\ell}.
\end{equation}

The effect of back reaction can be calculated (using the
two-dimensional Stefan-Boltzmann equation) as in the Appendix, and
leads to a rate of mass evaporation of the black hole.  The black hole can be treated 
as a black body of radius $r_H$ with a surface {\it length}, $2\pi r_H$, 
where 
\begin{equation}
\frac{dE}{dt} = - \zeta (3) \left (\frac{27}{32}\right ) \frac{\hbar^2 c^2}{\ell^2}
\left (\frac{G^{(2)} M}{c^2} \right ),
\end{equation}
and the Schwarzschild mass decreases $dM/dt = (dE/dt)/c^2$, and, since 
$\zeta (3) (27/32)$ $\approx 1$, we have
\begin{equation}
\frac{dM}{dt} \approx -\frac{\hbar}{\ell^2} \left (\frac{G^{(2)} M}{c^2} \right )^2 =
-\frac{M_{\rm Pl} c}{L_{\rm Pl}}\left (\frac{L_{\rm Pl}}{\ell} \right )^2 
\left (\frac{M}{M_{\rm Pl}} \right )^2,
\end{equation}
or, in our units,
\begin{equation}
\frac{dM}{dt} = -\frac{M^2}{\ell^2}.\label{dmdt}
\end{equation}
this can be solved for $M(t)$ as
\begin{equation}
M(t) = \frac{M(0)}{\frac{M(0)}{\ell^2} t + 1},\label{m(t)}
\end{equation}
assuming the $M(t)$ at $t = 0$ is $M(0)$.

With these elements we can consider the complete collapse scenario.  We
have to ask whether the classical shell represented by $<\hat R>(\tau)$ can fall
below the classical horizon radius $r_H = \ell \sqrt{M(0)}$ and return through 
the smaller horizon at $r_H = \ell \sqrt{M(t)}$ at some later, finite, $t$-time.
Naively, one might think that the falling horizon radius will meet the rising
shell that has passed through its minimum radius and is rebounding, ``sooner''
than it would have in the static case.  If we were to consider that we have 
been using some sort of ``Newtonian'' time, which is unique, this would 
always happen.  However, we are trying to model relativity, where $t$
should be observer time at large $r$ which is not the same as $\tau$,
the proper time on the shell.

Classically, we can find a relation between $t$ and $\tau$ by using the 
metric (\ref{metrc}) with $\theta = \theta_0 = $ constant., and $r = R(\tau)$,
\begin{equation}
-d\tau^2 = -\left (-M + \frac{R^2}{\ell^2} \right ) dt^2 + \frac{\dot R^2}
{-M + \frac{R^2}{\ell^2}} d\tau^2,
\end{equation}
or
\begin{equation}
\frac{dt}{d\tau} = \frac{\sqrt{\dot R^2 + R^2/\ell^2 - M}}{\left |
-M + \frac{R^2}{\ell^2} \right |}.\label{dtdtau}
\end{equation}
This expression has the problem that the denominator is zero whenever
$R = \ell \sqrt{M}$, so when the shell crosses the horizon (either descending 
or ascending), $dt/d\tau$
becomes singular, the origin of the fact that $t \rightarrow +\infty$ before
the rebounding shell can exit the horizon.

We want to use Eq. (\ref{dtdtau}) modified by quantum considerations with $M = M(t)$,
$M(t)$ given by (\ref{m(t)}).  Since $\dot R^2 + R^2/\ell^2 = 2E$, we will take this $2E$
to be $(2n + 1)/\ell$.  We now want to use $R(\tau) \rightarrow <\hat R>(\tau)$.
Unfortunately, the denominator of (\ref{dtdtau}) still becomes zero at $<\hat R>(\tau) =
\ell \sqrt{M(t)}$, and the relation between $t$ and $\tau$ is still singular at
this point.  Arguing that the quantum uncertainty in the shell position
makes the exact position $<\hat R>(\tau)$ unacceptable in our calculation
we can try to model this uncertainty by using $<\hat R^2> [M(t), \tau]$ in
our expression for $dt/t\tau$.  Finally, we will try to solve the simple
differential equation
\begin{equation}
\frac{dt}{d\tau} = \frac{\sqrt{\frac{2n + 1}{\ell} - M(t)}}{\left |-M(t) +
\frac{<\hat R^2 >[M(t), \tau]}{\ell^2}\right |},\label{dt2}
\end{equation}
with $M(t)$ given by (\ref{m(t)}).  This equation is a nonlinear, first-order differential
equation that must be solved numerically.

Our complete collapse scenario will now be:\\

1) Collapse from $R = \lambda \ell \sqrt{M(0)}$ to the horizon formation
point $R = \ell \sqrt{M(0)}$, with essentially no Hawking radiation, since 
the horizon is necessary for our calculation of $M(t)$.\\

2) From horizon formation to the point where 
$<\hat R>(\tau)$ is again above the horizon, we solve (\ref{dt2}) for $t(\tau)$ and
$M[t(\tau)]$, taking $t = 0$ at the point where the ring crosses the
horizon.\\

3) Study $<\hat R>(\tau)$ to see if, on the rebound, it crosses the
horizon in finite $t$-time.\\

\noindent If this scenario occurs, we will have to discuss the meaning of
the ``quantum remnant'' we have.

\vspace{0.5cm}
\centerline{\scalebox{.35}{\includegraphics{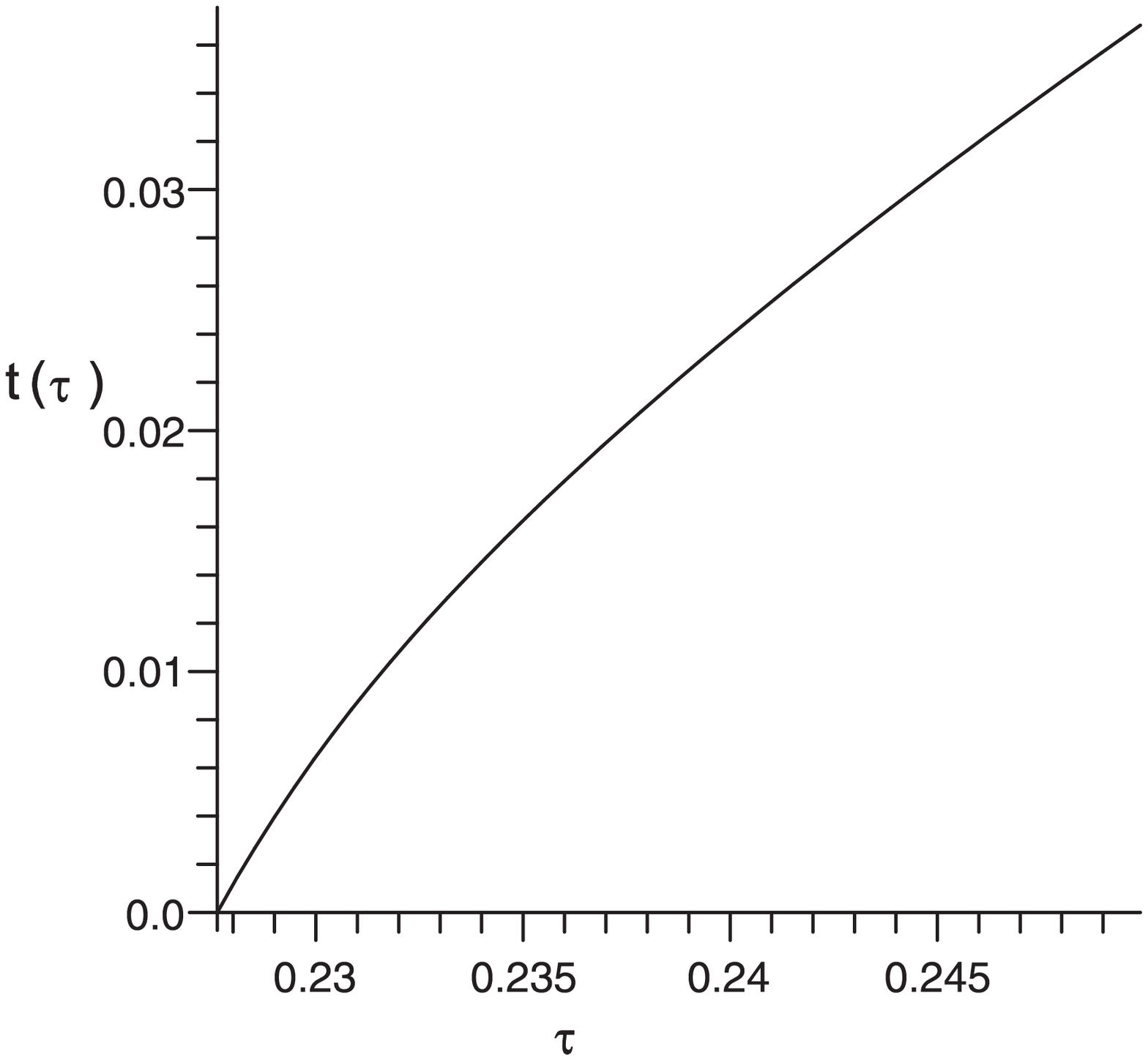}}}
\captionof{figure}{\small For $M(0) = 5$, the time $t$ at large $r$ as a function of $\tau$, the proper time on the shell.}

We will now take $M(0) = 5$ and $n = 3$ ($n$ must be odd), a somewhat
arbitrary choice.  Note that for $n = 3$ the maximum $M(0)$ for (\ref{dtdtau})
to make sense is $\approx 38.9$.  For $\lambda = 3$, $\ell = 0.18$, we
can calculate $t(\tau)$ numerically.  This was done using Maple, from
$\tau_0$ where the shell falls below the 
horizon, to $\tau_1$ where it would
reemerge classically with $M(0) =$ constant $= 5$.  The answer is given in Figure 4. 
The use of $<\hat R^2>$ makes $t$ a regular function of $\tau$ in this entire 
region.  Figure 5 shows the horizon size, $\ell \sqrt{M(t)}$.

\vspace{1cm}
\centerline{\scalebox{.45}{\includegraphics{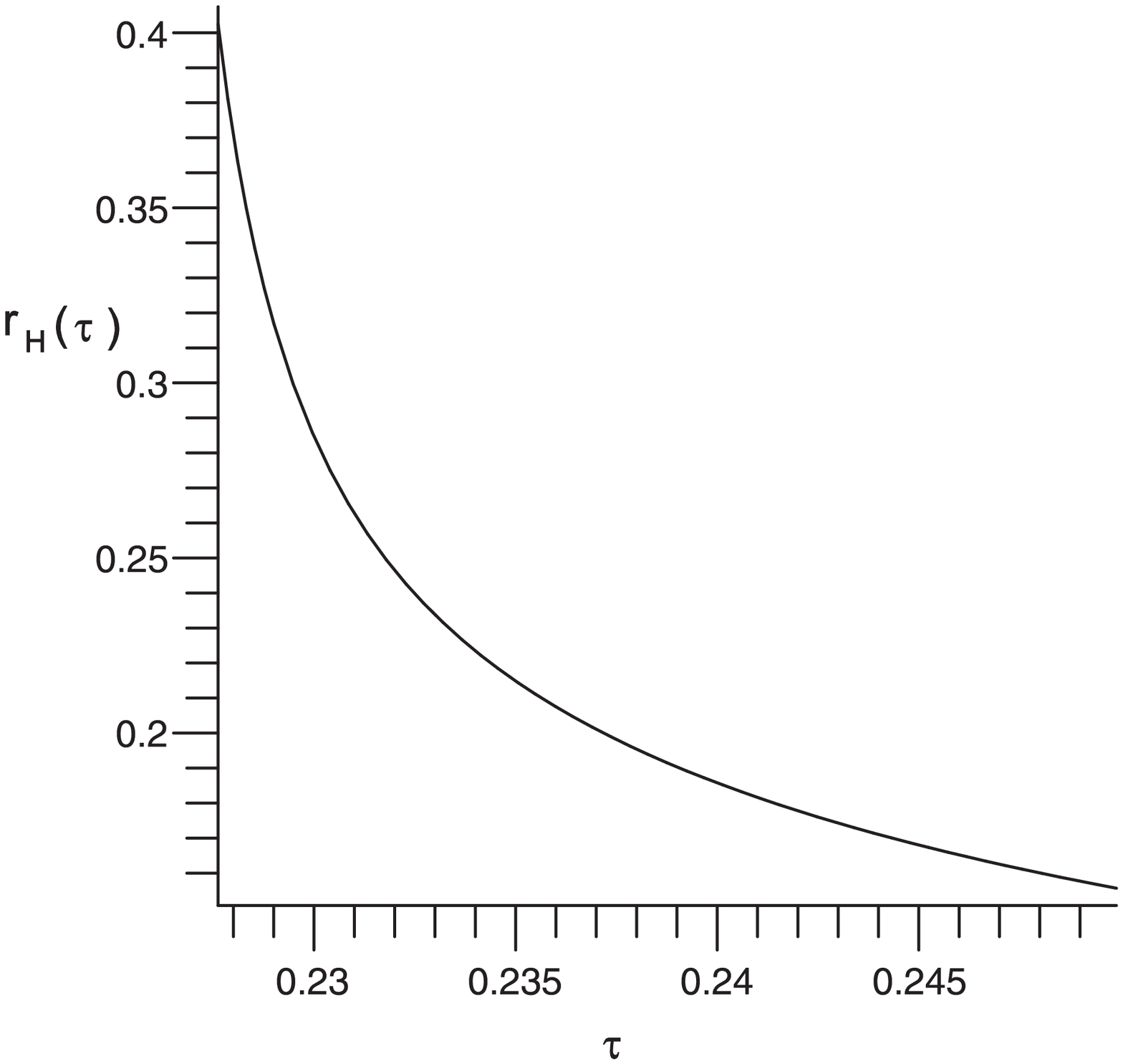}}}
\captionof{figure}{\small For $M(F0) = 5$, the horizon radius, $r_H = \ell \sqrt{M(\tau)}$ as a function of $\tau$.}
\vspace{0.5cm}
\vskip 15 pt

\noindent Figure 6
shows $<\hat R>(\tau)$ and the horizon size $\ell\sqrt{M[t(\tau)]}$ in this 
region.  We can see that $<\hat R>(\tau)$ would never cross the horizon, 
and we can interpret this to mean that Hawking radiation would change $M(t)$
so quickly that no horizon would form.

If we now take $M(0) = 10$, we can see from Figure 7 that the form of $t(\tau)$
does not change very much.  Figure 8 shows $\ell \sqrt{M[t(\tau)]}$, and
Figure 9 shows both $<\hat R>(\tau)$ and $\ell \sqrt{M[t(\tau)]}$.  
In this case, $<\hat R>(\tau)$ still does not fall below the horizon, but
a more detailed study of $<\hat R>(\tau)$ near the initial value of $\tau$
shows that the difference between $<\hat R>(\tau)$ and $r_H (\tau)$ is only
of the order of $10^{-5}$, so the shell grazes $r_H(\tau)$. 

\vspace{1cm}
\centerline{\scalebox{.4}{\includegraphics{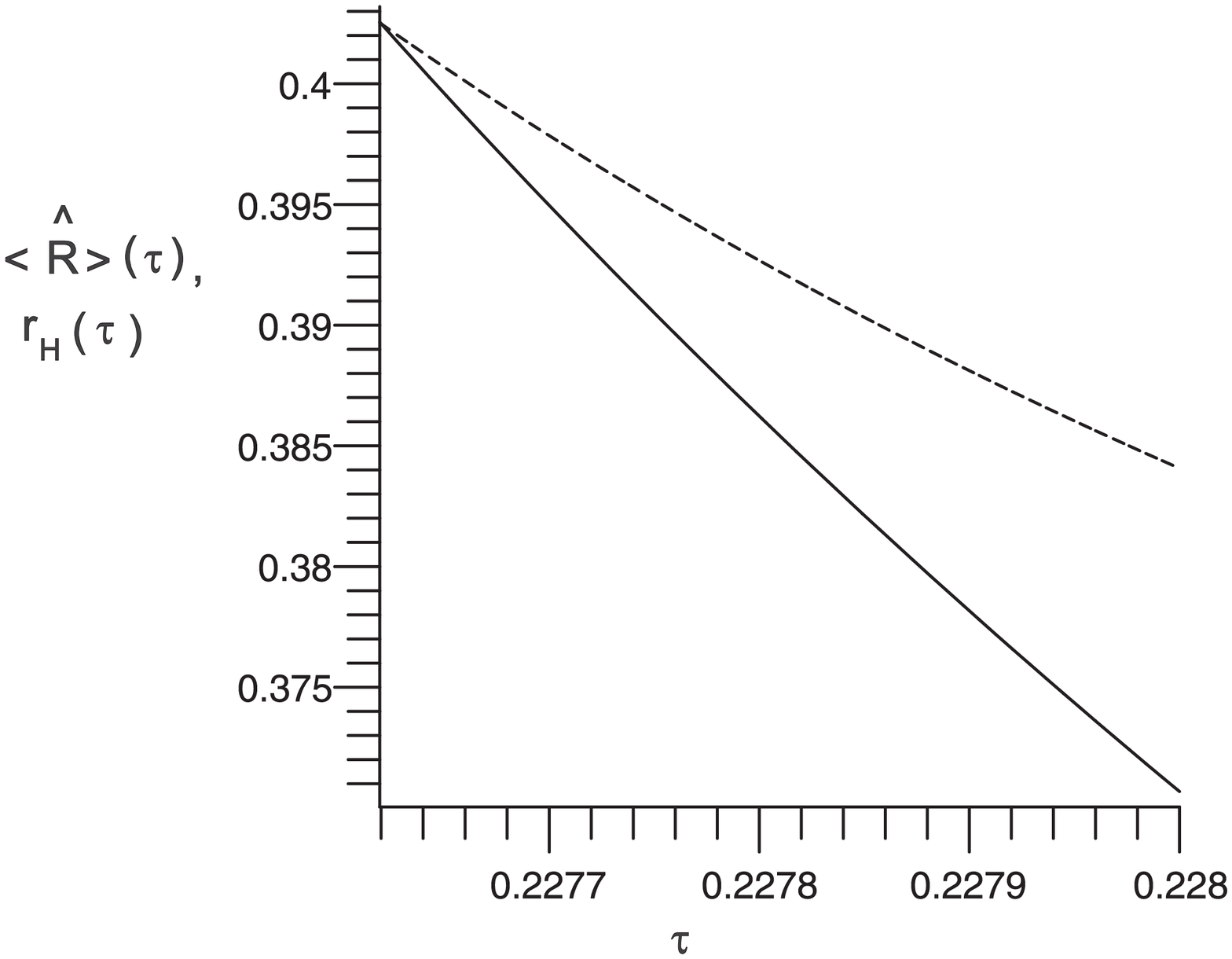}}}
\captionof{figure}{\small For $M(0) = 5$, the horizon radius $r_H$ (solid line), and the expectation value of the shell
radius $<\hat R>[M\{t(\tau)\}]$ (dashed line) as functions of $\tau$.}
\vspace{1cm}
\centerline{\scalebox{.4}{\includegraphics{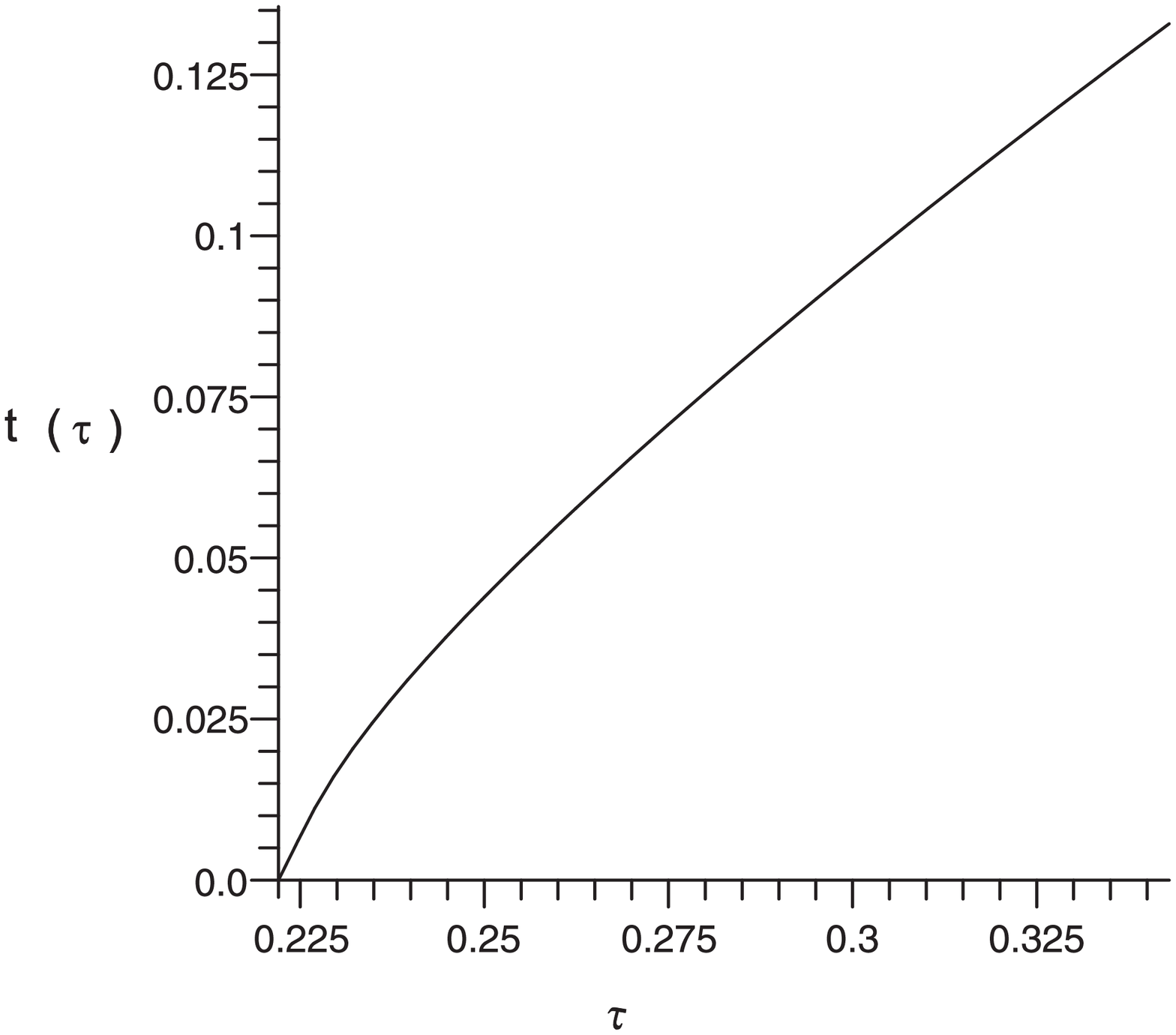}}}
\captionof{figure}{\small For $M(0) = 10$, the time $t$ at large $r$ as a function of $\tau$, the proper time on the shell.}
\vspace{0.5cm}
\centerline{\scalebox{.4}{\includegraphics{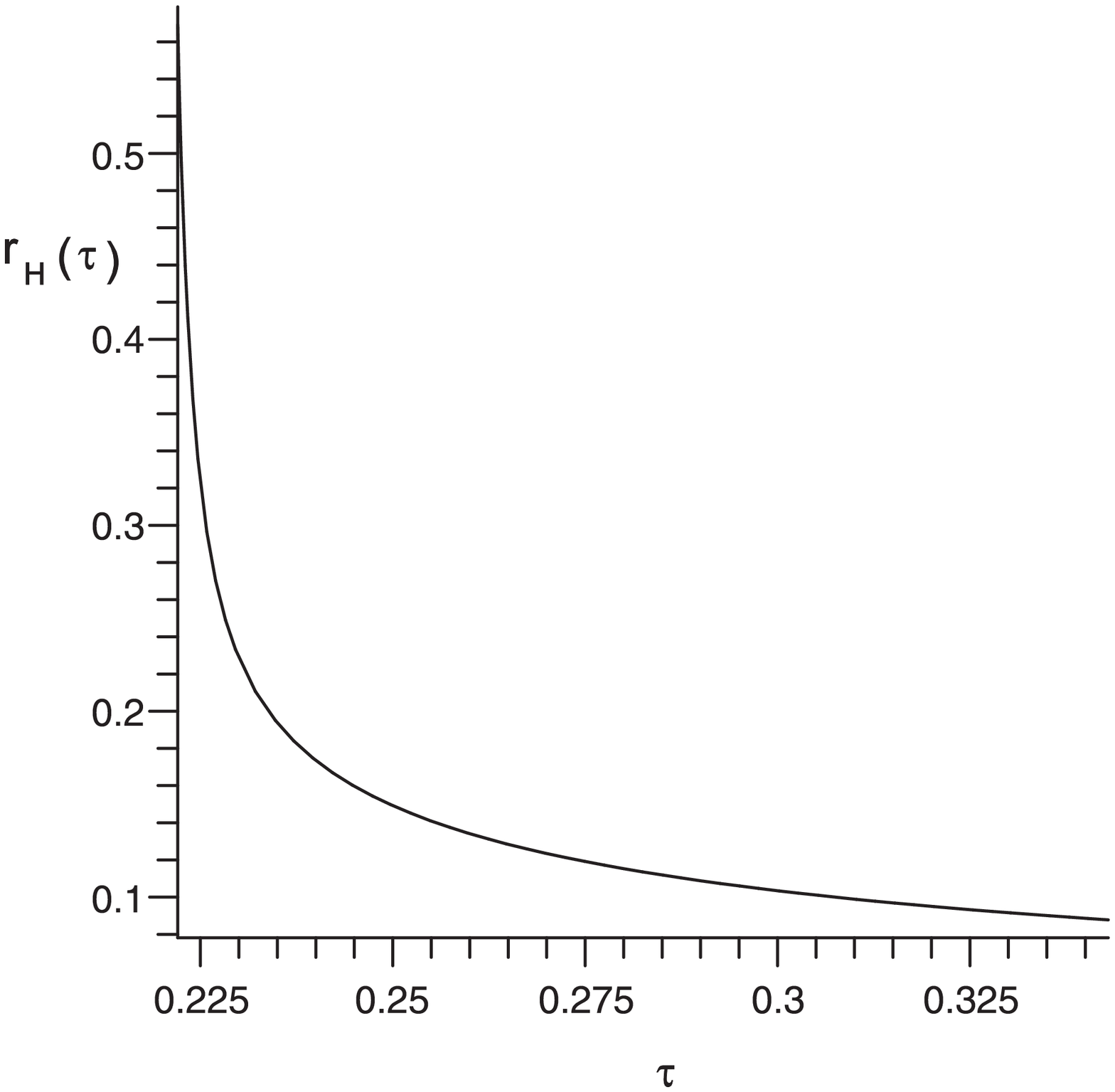}}}
\captionof{figure}{\small For $M(0) = 5$, the horizon radius $r_H$ (solid line), and the expectation value of the shell
radius $<\hat R>[M\{t(\tau)\}]$ (dashed line) as functions of $\tau$.}
\vspace{0.5cm}
\centerline{\scalebox{.4}{\includegraphics{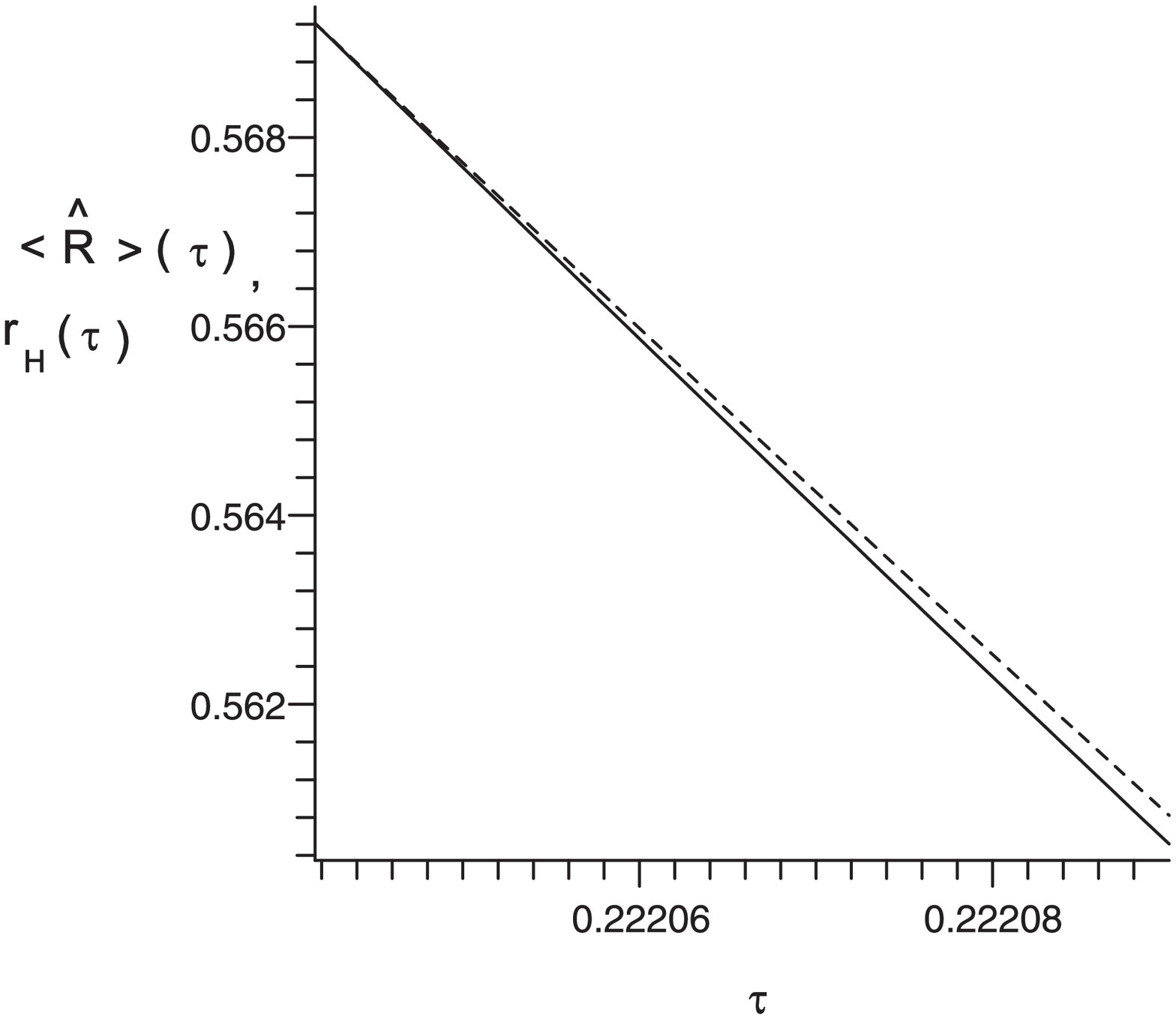}}}
\captionof{figure}{\small For $M(0) = 10$, the horizon radius $r_H$ (solid line), and the expectation value of the shell
radius $<\hat R>[M\{t(\tau)\}]$ (dashed line) as functions of $\tau$.}
\vspace{0.5cm}

If we now take $M(0) = 20$, we can see from Fig. 10 that the form of $t(\tau)$
still does not change very much.  Figure 11 shows $\ell \sqrt{M[t(\tau)]}$, and
Figure 12 shows both $<\hat R>(\tau)$ and $\ell \sqrt{M[t(\tau)]}$ from
$\tau \approx 0.22139$. 

\vspace{0.25cm}
\centerline{\scalebox{.4}{\includegraphics{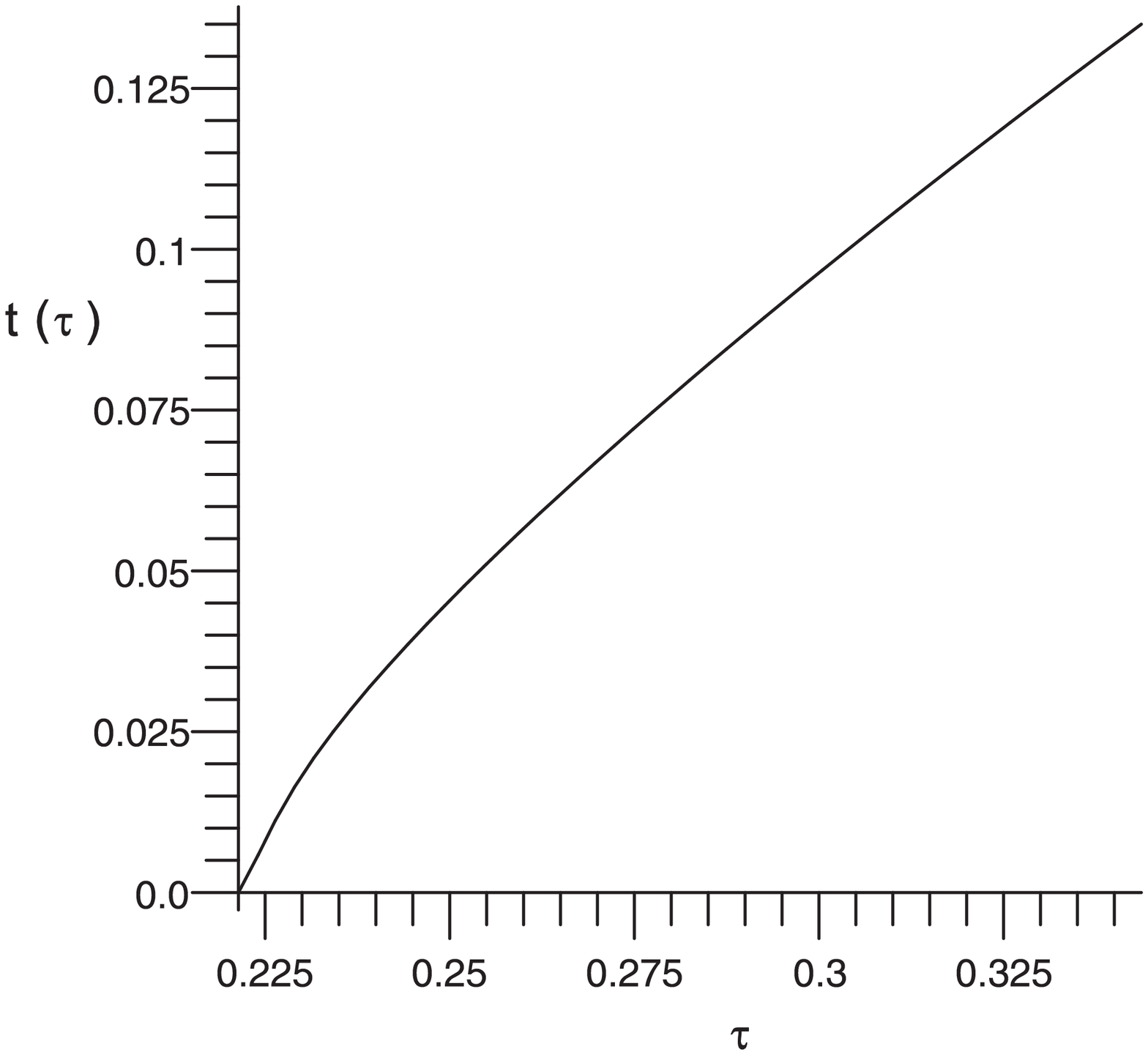}}}
\captionof{figure}{\small For $M(0) = 20$, the time $t$ at large $r$ as a function of $\tau$, the proper time on the shell.}

\centerline{\scalebox{.4}{\includegraphics{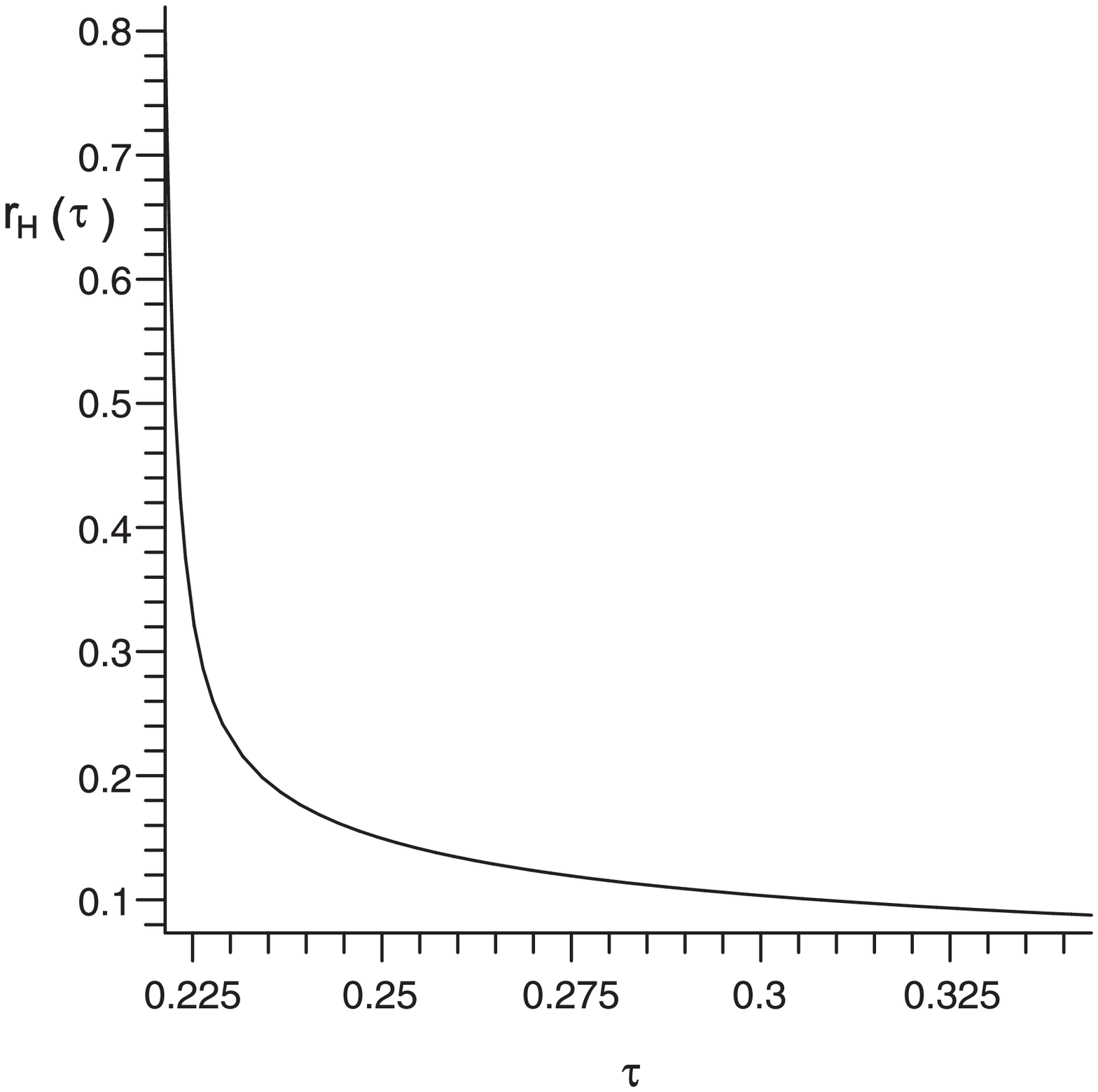}}}
\captionof{figure}{\small For $M(0) = 20$, the horizon radius, $r_H = \ell \sqrt{M(\tau)}$ as a function of $\tau$.}

\centerline{\scalebox{.4}{\includegraphics{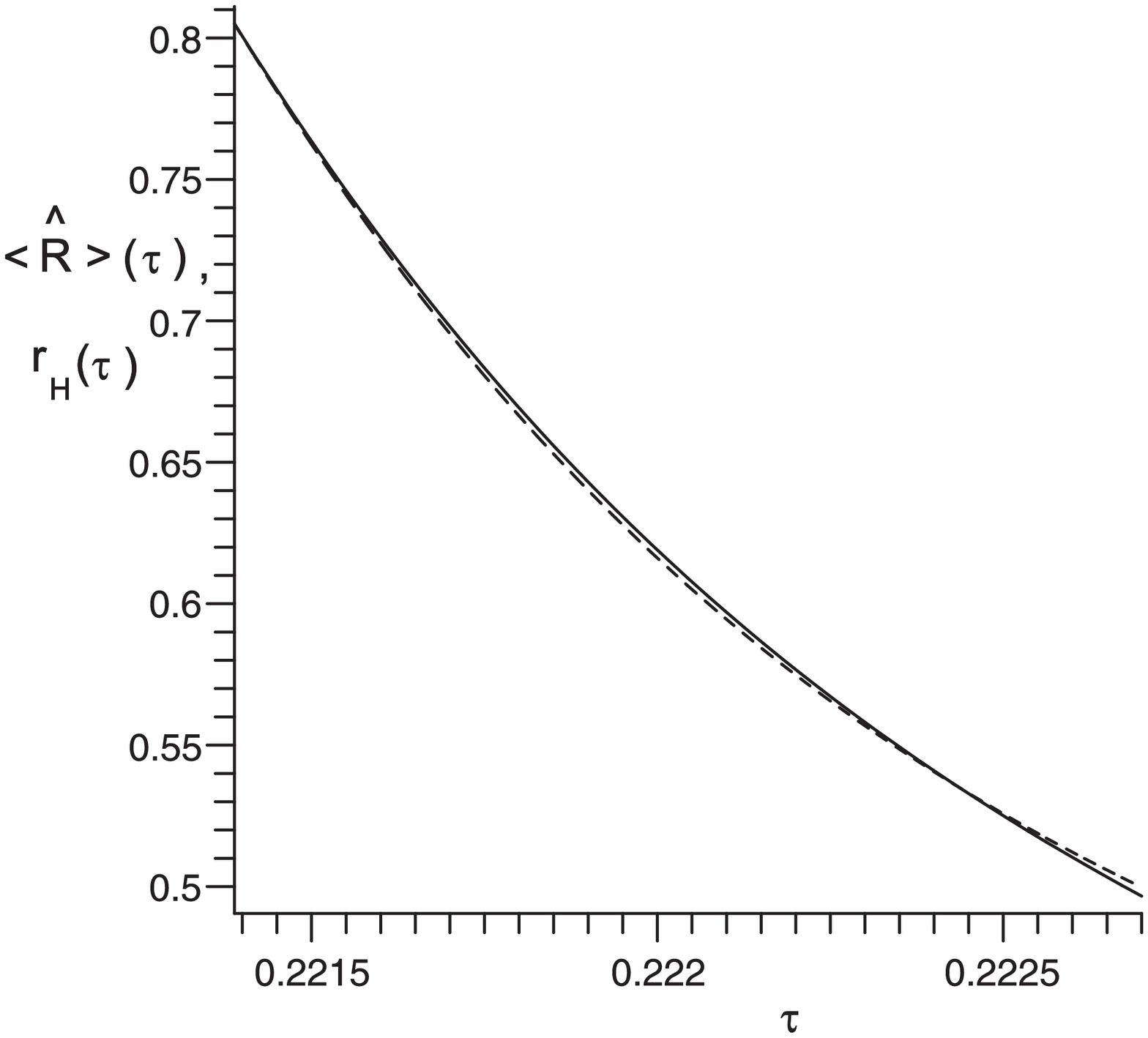}}}
\captionof{figure}{\small For $M(0) = 20$, the horizon radius $r_H$ (solid line), and the expectation value of the shell
radius $<\hat R>[M\{t(\tau)\}]$ (dashed line) as functions of $\tau$.}
\vspace{20pt}

\noindent Figure 13 shows ten times the difference between $<\hat R>[M\{t(\tau)\}]$ and $r_H(\tau)$,
making the dip below $r_H$ more obvious.

\vspace{0.5cm}
\centerline{\scalebox{.35}{\includegraphics{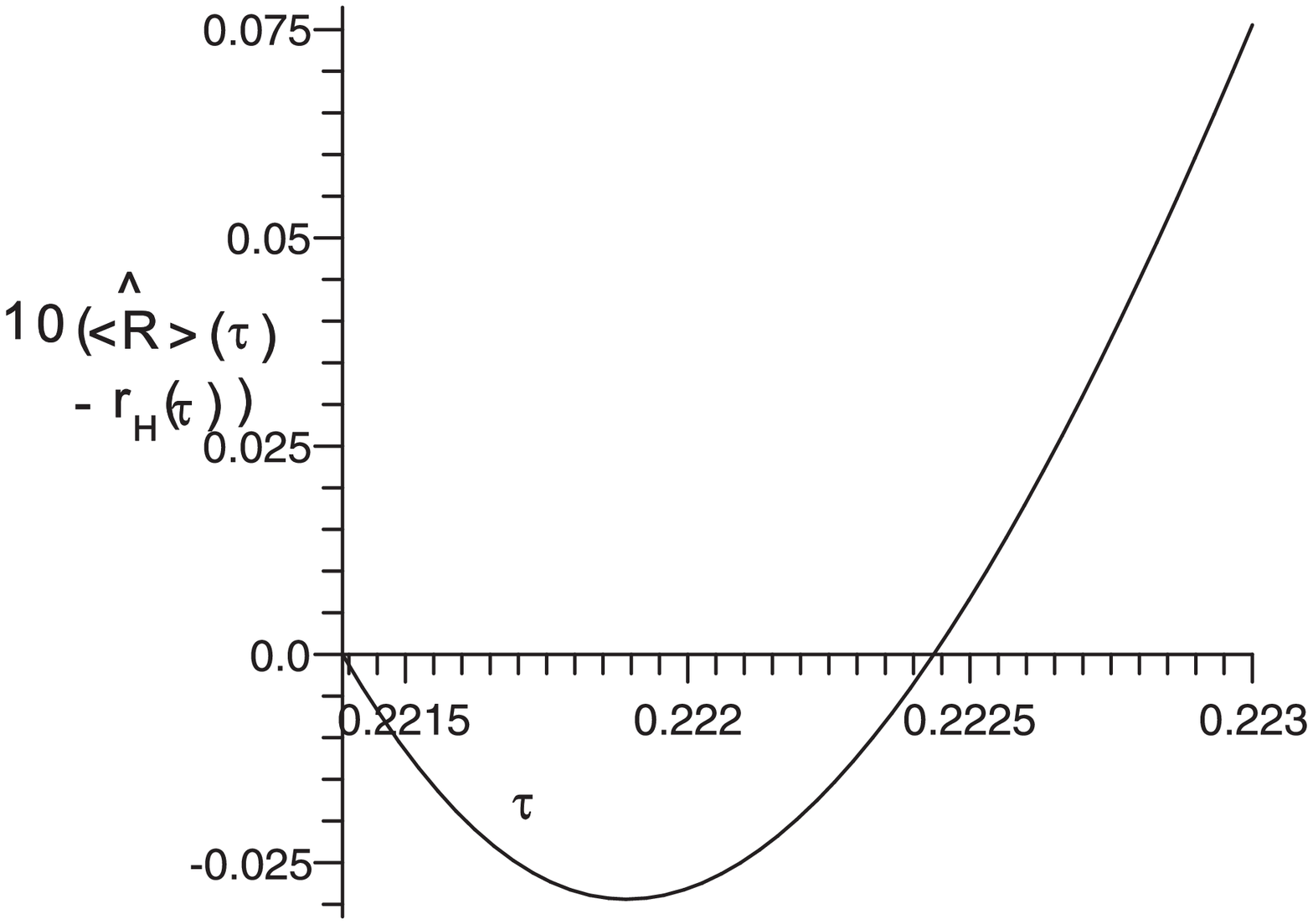}}}
\captionof{figure}{\small For $M(0) = 20$, ten times $<\hat R>(\tau) - r_H(\tau)$, showing the dip of the expectation value
below $r_H$ and its return above $r_H$.}
\vspace{0.5cm}

We can see that the shell falls below the horizon at $\tau \approx \tau_0$, $\approx 0.22139$ and
reemerges  at $\tau \approx 0.22241$, at which point Hawking radiation ceases.  
Since $t$ is a regular function of $\tau$ in this region, $t$ at reemergence
is finite.  We can interpret this scenario as loosely defining a ``dynamic remnant.''
Many calculations in 3 + 1 gravity of Hawking radiation back reaction assume
an eternal, static or stationary black hole, and that any remnant would be some sort 
of static or stationary construct.  In our case, the fact that our quantum shell
``rebounds,'' seems to mean that our ``remnant'' would be an expanding quantum shell.

The important point here is that there is mass loss during the Hawking radiation
epoch is $\approx 11$, or 55\%.  For $M(0) = 30$, the behavior is
similar, with mass losses of $\approx 20$, or 67\% but the loss is not 100\%.
As $M(0) \rightarrow \infty$, the mass loss grows, but it should never be 100\%.
In the 3 + 1 case, since $dM/dt$ decreases with growing $M$, we might expect the mass
loss to decrease.

Our scenario is the collapse below a horizon, mass loss due to Hawking radiation, and
the reappearance of the shell in expansion, but with a reduced mass.  We can consider
this a ``dynamic remnant.''  There are a number of problems with this calculation beyond
its rough nature.  One is that we have assumed in our Hawking radiation calculation that
the Compton wavelength, $\lambda_c$, is much smaller that $r_H$.  However, our units show
that $M$ is of the order of a few Planck masses, so $\lambda_c$ for any reasonable
particle is very much larger than $r_H$.  We would have to drastically modify our
Hawking radiation approximation in this case.  This problem reflects the well-known
difficulty of calculating Hawking radiation for particle masses above the horizon
radius (see \cite{koch} and references therein).

Another problem is that we can still have information loss.  The existence of a remnant
does not necessarily solve all problems.  There is always the necessity for an
infinite number of states which allows for the unbounded information content inherited
from the original state. 

\section{Conclusions and suggestions for further research}  

While the above results are suggestive, there are several caveats, and
one should be careful in interpreting the results.  One problem is that
Eq. (\ref{dmdt}) has the rate of mass loss {\it increasing} as $M$ increases, while
in 3 + 1 gravity it {\it decreases}, so even if we accept the results, they
will be numerically quite different from those in 3 + 1 gravity.  Of course, 
our results are, at best, crude, due to the drastic simplifications we have 
made.  Perhaps the most sensitive choices we have made are in the equation
for $dt/d\tau$.  If we were to use the approximation of the classical $T^{ij}$
constructed from $\psi^* \psi$, some of these difficulties could be sidestepped, 
but there will still be a difficult problem in that the proper time $\tau$ is 
valid only for {\it one} shell, and we would somehow have to change the 
problem to account for different proper times on different shells.  One advantage
we have in our simplified calculation is that we can argue that our proper time
is valid for the {\it one} shell we have.

Another time problem is that we are using (essentially) Schwarzschild time for
$t$, while a coordinate system such as Kruskal might be a better
choice.  However, the coordinate change from Schwarzschild contains $M$,
and a changing $M$ makes the coordinate change difficult.  In the classical
$T^{ij}$ problem, we would have to {\it invent} the equivalent of Kruskal
coordinates.

The one advantage of 2 + 1 gravity is that the quantum collapse problem is
exactly solvable, but numerical work in 3 + 1 would be more believable.

However, we can make several suggestions for further research:
\vskip 20 pt

1) Do the $T^{ij}$ problem, with due attention to the difficulties with
proper time.\\

2) Return to 3 + 1 gravity and try to do a similar simplified calculation
to the one attempted here.\\

3)Redo the calculation with due care given to the fact that, in general
$\lambda_c >> r_H$.
\vskip 20 pt

\noindent Each of these would require numerical solutions.
 
\section*{Appendix}

We want to do the same simplified calculation of Hawking radiation
energy that we did in Sec. 4 for the 2 + 1 black hole.  If we use 
Fig. 3, taking the circles to be spheres, we have a thin shell between
$r_H$ and $r_H + \lambda_c$ where we can have the production of virtual 
particles with mass $m$, with one of the pair falling into the hole,
and the other becoming real with outward radial velocity $c$.  The
(constant) Newtonian energy of the particle, if produced at $r =
R_H + \lambda_c/2$, is
\begin{equation}
E = \frac{mc^2}{2} - \frac{GMm}{r_H + \lambda_c/2},
\end{equation}
and if $\lambda_c << r_H$,
\begin{equation}
E \approx \frac{mc^2}{2} - \frac{GMm}{r_H}\left (1 - \frac{\lambda_c}{2r_H}\right ),
\end{equation}
\begin{equation}
= \frac{mc^2}{2} - \frac{mc^2}{2} + \frac{\hbar c^3}{8GM} = \frac{\hbar c^3}{8GM}.
\end{equation}
The exact average energy of Hawking radiation particles at infinity is
\begin{equation}
\left( \frac{\Gamma (4) \zeta (4)}{\pi \Gamma(3) \zeta (3)}\right ) \frac{\hbar c^3}{8GM},
\end{equation}
which is within 14\% of our simplified result.

We can now define a Hawking temperature
\begin{equation}
T_H = \frac{\hbar c^3}{8GM k_B},
\end{equation}
and treating the black hole as a black body with radius $r_H$, we have an
energy loss due to Hawking radiation as ($\sigma$ the Stefan-Boltzmann constant)
\begin{equation}
\frac{dE}{dt} = -\sigma (4\pi r_H^2) T_H^4,
\end{equation}
\begin{equation}
= -\frac{\pi^3}{15360}\left (\frac{\hbar c^6}{G^2}\right ) \left(\frac{1}{M^2}\right).
\end{equation}
The mass loss equation is
\begin{equation}
\frac{dM}{dt} = -\frac{\pi^3}{15360}\left (\frac{\hbar c^4}{G^2}\right ) \left(\frac{1}{M^2}\right).
\end{equation}
Note that $dM/dt$ {\it decreases} with increasing $M$, rather than growing
as in the 2 + 1 case.  Also, $T_H$ is {\it independent} of $m$, so, formally, we
can take $m \rightarrow 0$ to assume the production of massless particles.

Notice that we have assumed $r_H << \lambda_c$, and if we wanted to use
our simplified calculation for Planck-sized black holes, we would have to 
change our approximation.

\end{document}